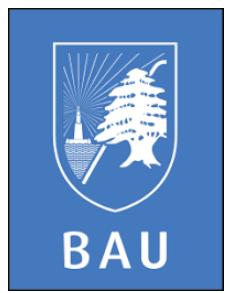

SENIOR PROJECT REPORT

# Non-existence of Fermionic Bound States in Schwarzschild Geometry

Student: Mohamad Fayez Shehadeh

Supervisors:

Dr. Mohamad Salem Badawi  Associate Professor, Physics Department

Dr. Toufic El Arwadi  Associate Professor, Mathematics and Computer Science Department

Dirac equation in Newman-Penrose formalism is comprehensively introduced to prove the non-existence of bound states for $\omega < m_e$ in Schwarzschild geometry. The proof by contradiction is drawn in the context of finding polynomial solutions for the generalized Heun equation, thus the solutions are resonant states decaying over time. This work is divided into five parts. The first part is about Dirac equation and the circumstances in which this equation emerged. Followed by two brief subsections on the mathematical substratum of Dirac equation and its solution, i.e., Dirac algebra and Dirac spinor respectively. For the sake of elaborating the mathematical tools used in this subject, the second part consists of a brief motivation for the general theory of relativity and its mathematical methods. The third part is dedicated to treating Dirac equation in a more advanced mathematical framework, the Newman-Penrose formalism, which is a special case of the tetrad formalism of general relativity. The proof is elaborated in part 4, followed by conclusions in the fifth part.

# Table of Contents

*"What is a spinor?*

*I spent most of my life studying spinors in one way or another, and I don't know. Only God knows… maybe Dirac knows, but he's no longer with us"*

SIR MICHEAL ATIYAH

## I-Dirac Equation

### 1-From Schrodinger's and Klein-Gordon to Dirac's:

#### 1-1-Schrödinger Equation:

Adapting the natural units $\hbar = c = 1$ and the canonical commutator is $[\hat{p}_i, \hat{x}_j] = -i\delta_{ij}$ with coordinate representation: $\vec{p} = -i\vec{\nabla}$ and $\hat{H}\psi(\vec{x}, t) = i\frac{\partial \psi(\vec{x},t)}{\partial t}$. The Hamiltonian for spinless particle in Schrödinger theory is $\hat{H} = \frac{\vec{p}^2}{2m} + \hat{V}$, so the Schrödinger equation reads

$$\left[\frac{-1}{2m}\vec{\nabla}^2 + \hat{V}(\vec{x}, t)\right]\psi(\vec{x}, t) = i\frac{\partial \psi(\vec{x}, t)}{\partial t}$$

The resulting probability density and probability current are:

$$\rho = \psi^*\psi = |\psi|^2 \geq 0$$

$$\vec{j} = \frac{1}{2im}\left[\psi^*(\vec{\nabla}\psi) - (\vec{\nabla}\psi^*)\psi\right]$$

This yields to the continuity equation

$$\frac{\partial \rho}{\partial t} + \vec{\nabla}.\vec{j} = 0$$

The free particle solution for Schrödinger equation is $\psi(x, t) = u(x)e^{-iEt}$ with $H_0 u = Eu$ where $H_0$ corresponds to V=0.

Schrodinger equation is non-relativistic wave equation. Several physicists tried to get a relativistic version of this equation, mainly Oskar Klein and Walter Gordon in 1926.

#### 1-2-Klein-Gordon Equation:

The non-relativistic Schrodinger equation may be put in correspondence with the non-relativistic energy-momentum relation $E = \frac{p^2}{2m}$ by means of the operator replacements:

$$E \rightarrow i\frac{\partial}{\partial t}, \quad p \rightarrow -i\nabla$$

Klein and Gordon attempted to build a relativistic quantum mechanics on the relativistic energy-momentum relation $E^2 = p^2 + m^2$. By the operator replacements, we get

$$-\frac{\partial^2 \varphi}{\partial t^2} = (-\nabla^2 + m^2)\varphi \qquad (1)$$

This is the Klein-Gordon equation. $\varphi(x, t)$ is a scalar wavefunction, meaning that it is appropriate for describing spinless particles, i.e., bosons.

Using the d'Alembertian1 $\Box \equiv \partial_\mu \partial^\mu = \frac{\partial^2}{\partial t^2} - \nabla^2$ , the Klein-Gordon equation reads

$$(\Box + m^2)\varphi(x, t) = 0$$

The plane-wave solution of this equation is of the form

$$\varphi(x, t) = Ne^{-iEt + ip.x} = Ne^{-ipx}$$

Where2 $px = p_\mu x^\mu = Et - p.x$. Energy $E$ is related to $p$ by $E^2 = p^2 + m^2$, so for a given 3-momentum $p$, there are two solutions for the energy $E = \pm(p^2 + m^2)^{1/2}$. This implies that a particle can decay to an infinite number of negative states and as it decays and loses energy, its energy becomes more and more negative in a process that implies that its momentum grows infinitely.

Starting from equation (1) and multiplying by $\varphi^*$ then subtracting $\varphi\varphi^*$, we get the continuity equation

$$\frac{\partial \rho}{\partial t} + \vec{\nabla}.\vec{j} = 0 \qquad (2)$$

where,

$$\rho = i\left[ \varphi^* \frac{\partial \varphi}{\partial t} - \left(\frac{\partial \varphi^*}{\partial t}\right)\varphi \right]$$

$$j = \frac{1}{i}[\varphi^* \nabla \varphi - (\nabla \varphi^*)\varphi]$$

In 4-vector notation, equation (2) reads

$$\partial_\mu j^\mu = 0$$

with $j^\mu = (\rho, j) = i[\varphi^* \partial^\mu \varphi - (\partial^\mu \varphi^*)\varphi]$. The probability density $\rho$ is not constrained to be positive. For the plane-wave solution

$$\varphi = Ne^{-iEt + ip.x}$$

the probability density is $\rho = 2|N|^2 E$. The negative energy solutions are associated with negative probability.[1]

### 1-3-Dirac Equation:

In this section, we consider the Dirac equation as an evolution equation in a suitable Hilbert space whose vectors are related to the physical states via a statistical interpretation.[4]

---

1 Einstein summation rule is adapted

2 Note that $p_\mu = (p_0 = E, p_1, p_2, p_3)$ and $x_\mu = (x_0 = t, x_1, x_2, x_3)$

In 1927, Dirac devised a relativistic wave equation linear in $\partial/\partial t$ and $\nabla$ to avoid these problems[2] and include external electromagnetic fields in an appropriate relativistically invariant way[4]. He succeeded in overcoming the problem of negative probability with an unexpected cumshaw that his equation describes a spin-½ particle. Later, Stuckelberg and Feynman interpreted the negative energy solutions as a particle propagates backward in time, or equivalently, a positive energy antiparticle propagating forward in time.[2]

The Dirac Hamiltonian which is linear in the energy, mass, and momentum, is of the form

$$H = \vec{\alpha}.\vec{p} + \beta m = \alpha_i p_i + \beta m$$

where $H\Psi = E\Psi$. So, the Dirac equation is

$$i\frac{\partial \psi}{\partial t} = \left[-i\left(\alpha_1 \frac{\partial}{\partial x_1} + \alpha_2 \frac{\partial}{\partial x_2} + \alpha_3 \frac{\partial}{\partial x_3}\right) + \beta m\right]\psi = (-i\alpha.\nabla + \beta m)\psi \quad (3)$$

The task here is to find a solution satisfying the Klein-Gordon condition, which is Lorentz covariant and appropriate for relativistic mechanics, in addition to find conditions of $\alpha$ and $\beta$.

Squaring the operators on both sides of equation (3):

$$\left(i\frac{\partial}{\partial t}\right)^2 \psi = (-i\alpha.\nabla + \beta m)(-i\alpha.\nabla + \beta m)\psi$$

$$= -\sum_{i=1}^{3} \alpha_i^2 \frac{\partial^2 \psi}{(\partial x^i)^2} - \sum_{\substack{i,j=1 \\ i \neq j}}^{3} (\alpha_i \alpha_j + \alpha_j \alpha_i) \frac{\partial^2 \psi}{\partial x^i \partial x^j} - im \sum_{i=1}^{3} (\alpha_i \beta + \beta \alpha_i) \frac{\partial \psi}{\partial x^i} + \beta^2 m^2 \psi \quad (4)$$

In order for $\psi$ to satisfy equation (4), it requires that $\alpha$ and $\beta$ anticommute: $\begin{cases} \alpha_i \beta + \beta \alpha_i = 0 \\ \alpha_i \alpha_j + \alpha_j \alpha_i = 0 \end{cases}$ and $\alpha_i^2 + \beta^2 = 1$.

$\alpha$'s and $\beta$ are matrices since numbers does not anticommute. They are interpreted as matrices acting on a wavefunction, the physical states they represent have the same energy associated with a new degree of freedom: the spin. The lowest dimensionality matrices satisfying these conditions are 4×4 [2]. The choice of the four matrices ($\boldsymbol{\alpha}, \beta$) is not unique, the Dirac-Pauli representation is mostly used [2]:

$$\alpha = \begin{pmatrix} 0 & \sigma \\ \sigma & 0 \end{pmatrix}, \qquad \beta = \begin{pmatrix} I & 0 \\ 0 & -I \end{pmatrix}$$

Where $\sigma$ are the Pauli matrices:

$$\sigma_1 = \begin{pmatrix} 0 & 1 \\ 1 & 0 \end{pmatrix}, \qquad \sigma_2 = \begin{pmatrix} 0 & -i \\ i & 0 \end{pmatrix}, \qquad \sigma_3 = \begin{pmatrix} 1 & 0 \\ 0 & -1 \end{pmatrix}$$

We introduce Dirac $\gamma$-matrices defined by[3]

$$\gamma^\mu \equiv (\beta, \beta\boldsymbol{\alpha})$$

In Minkowski spacetime, Dirac equation for massive ½-spin particle is of the form

$$i\gamma^\mu \partial_\mu \psi - m\psi = 0 \qquad (5)$$

or by using Feynman 'slash' notation $\not{p} = \gamma^\mu p_\mu$ we get

$$(\not{p} - mI)\psi = 0 \qquad (6)$$

The corresponding solution is of the from $\psi = \begin{pmatrix} \psi_1 \\ \psi_2 \\ \psi_3 \\ \psi_4 \end{pmatrix}$ (7)

Equation (5) has the explicit form:

$$i\gamma^0 \partial_t \psi + i\gamma^1 \partial_x \psi + i\gamma^2 \partial_y \psi + i\gamma^3 \partial_z \psi - m\psi = 0 \qquad (8)$$

With equation (7), we get a set of coupled partial differential equations:

$$i\partial_t \psi_1 + i\partial_z \psi_3 + i\partial_x \psi_4 + \partial_y \psi_4 - m\psi_1 = 0$$

$$i\partial_t \psi_2 + i\partial_x \psi_3 - \partial_y \psi_3 - i\partial_z \psi_4 - m\psi_2 = 0$$

$$-i\partial_z \psi_1 - i\partial_x \psi_2 - \partial_y \psi_2 - i\partial_t \psi_3 - m\psi_3 = 0$$

$$-i\partial_x \psi_1 + \partial_y \psi_1 + i\partial_z \psi_2 - i\partial_t \psi_4 - m\psi_4 = 0$$

In Minkowski spacetime, the 4-components spinor[3] $\psi$ are of the form

$$\psi^{(+)} = u(p)e^{-ip_\mu x^\mu} \qquad (9)$$

and

$$\psi^{(-)} = v(p)e^{ip_\mu x^\mu} \qquad (10)$$

where $p_t = p^t \geq 0$ and $\psi^{(+)}$ and $\psi^{(-)}$ denote the positive and negative energy solutions, respectively. Substituting equations (9) and (10) in equation (6) we get

$$(\not{p} - mI)u(p) = 0 \qquad (11)$$

and

$$(\not{p} + mI)v(p) = 0 \qquad (12)$$

which are algebraic equations, it's a system of homogeneous equations for $u(p)$ and $v(p)$ having a not trivial solution only if $\det(\not{p} \pm mI) = 0$. This equation gives the condition

$$(p^2 - m^2)^2 = 0 \qquad (13)$$

where,

---

[3] cf. p. 12

$p^2 = (p^t)^2 - \vec{p}^2 \equiv E^2 - \vec{p}^2$. So, equation (13) can be written as

$$(p^2 - m^2) = \left( E - \sqrt{\vec{p}^2 + m^2} \right) \left( E + \sqrt{\vec{p}^2 + m^2} \right) = 0$$

The two positive energy spinors $u^{(1)}(p)$ and $u^{(2)}(p)$ satisfy equation (11), where the superscripts *(1)* and *(2)* refer to $S_z = +\frac{1}{2}$ and $S_z = -\frac{1}{2}$ respectively,

$$u^{(1)}(p) = N \begin{pmatrix} 1 \\ 0 \\ -\dfrac{p_z}{p_t + m} \\ \dfrac{-p_x + i p_y}{p_t + m} \end{pmatrix}, \qquad u^{(2)}(p) = N \begin{pmatrix} 1 \\ 0 \\ \dfrac{-p_x + i p_y}{p_t + m} \\ \dfrac{p_z}{p_t + m} \end{pmatrix}$$

Using equation (12),

$$v^{(1)}(p) = N \begin{pmatrix} -\dfrac{p_z}{p_t + m} \\ \dfrac{-p_x + i p_y}{p_t + m} \\ 1 \\ 0 \end{pmatrix}, \qquad v^{(2)}(p) = N \begin{pmatrix} \dfrac{-p_x + i p_y}{p_t + m} \\ \dfrac{p_z}{p_t + m} \\ 0 \\ 1 \end{pmatrix}$$

These four solutions are linearly independent. Then for the *x*-space, i.e., equation (9) and (10) we have:

$$\psi^{(+)(\alpha)}(x) = \begin{pmatrix} u_1^{(\alpha)}(p) \\ u_2^{(\alpha)}(p) \\ u_3^{(\alpha)}(p) \\ u_4^{(\alpha)}(p) \end{pmatrix} e^{-i p_\mu x^\mu}, \qquad \psi^{(-)(\alpha)}(x) = \begin{pmatrix} v_1^{(\alpha)}(p) \\ v_2^{(\alpha)}(p) \\ v_3^{(\alpha)}(p) \\ v_4^{(\alpha)}(p) \end{pmatrix} e^{i p_\mu x^\mu}$$

In short,

$$\psi^{(+)(\alpha)}(x) = u^{(\alpha)}(p) e^{-i p_\mu x^\mu}$$

$$\psi^{(-)(\alpha)}(x) = v(p) e^{i p_\mu x^\mu}$$

Consider the quantity[1] $\rho = \psi^* \psi$, so

$$\rho = \sum_{a=1}^{4} |\psi_a|^2 > 0$$

$\rho$ is a scalar density explicitly positive-definite. The other element for a conservative law emerging from Dirac equation is the probability current. By equation (3) and its Hermitian conjugate, we find that the conservative law is of the form

$$\frac{\partial \rho}{\partial t} + \vec{\nabla}.\vec{j} = 0$$

and $j = \psi^* \alpha \psi$, which represent a 3-vector. By this, Dirac solved the negative probability problem.

## 2- Dirac Algebra:

There are two main approaches to the study of geometry, referred to as 'algebraic' and 'synthetic' traditions. The algebraic tradition deals with the manipulation of the components of vectors. That naturally leads to the subject of tensors and emphasizes how coordinates transform under the change of frame. The synthetic approach treats vectors as single, abstract entities, and manipulates them directly. Geometric algebra follows in tradition.[5] The foundations of geometric algebra were laid by Hamilton and Grassmann in the 19th century. Clifford unified their work, and the result is the Clifford algebra. It was rediscovered by Pauli and Dirac in the quantum theory of spin. In its fundamental level, geometric algebra is a mathematical language for direct encoding of geometric primitives such as points, lines, planes, volumes, etc.[6]

### 2-1-Dirac Algebra:

The original equation used by Dirac to describe a free electron is the matrix equation[7]

$$\left(i\gamma_0 \frac{\partial}{\partial t} - \gamma_1 \frac{\partial}{\partial x_1} - \gamma_2 \frac{\partial}{\partial x_2} - \gamma_3 \frac{\partial}{\partial x_3}\right)\psi = m\psi$$

The gamma matrices $\gamma_0, \gamma_1, \gamma_2, \gamma_3$ are 4×4 matrices operating on the wave function $\psi$, the components of which describe the four charge and spin states of the electron. Each gamma matrix mutually anticommute and has a square of $+I$

$$\gamma_\mu \gamma_\nu = -\gamma_\nu \gamma_\mu, \qquad \mu \neq \nu \quad (13)$$

$$\gamma_\mu{}^2 = I \qquad (14)$$

By taking the Hermitian conjugate of Dirac equation, we find that gamma matrices are Hermitian (in the standard Pauli-Dirac representation)

$$\gamma_\mu{}^\dagger = \gamma_\mu$$

Besides $\gamma_\mu$ and the identity, additional matrices are used to perform rotations, Lorentz transformations, the parity and time reversal operations, and so forth. The additional matrices, apart from $\gamma_\mu$ and the identity, are a set of 16 matrices forming a representation of Dirac algebra. In the full Dirac algebra, gamma matrices either commute or anticommute, the whole structure of Dirac algebra is determined by these relations.

Because of the conditions imposed by equation (13) and (14), the Dirac algebra is a Clifford algebra. In a Clifford algebra, one starts with certain number of mutually anticommuting elements and multiplies them together to obtain the remaining elements, the Dirac algebra has four initial elements. Any product of gammas with repeated indices can be simplified, e.g., $\gamma_1 \gamma_2 \gamma_1 = -\gamma_2 \gamma_1 \gamma_1 = -\gamma_2$. There are six possible bilinear products:

$$\gamma_{12} = -i\gamma_1\gamma_2, \qquad \gamma_{23} = -i\gamma_2\gamma_3, \qquad \gamma_{31} = -i\gamma_3\gamma_1, \qquad \gamma_{10} = -i\gamma_1\gamma_0, \qquad \gamma_{20} = -i\gamma_2\gamma_0,$$

$$\gamma_{30} = -i\gamma_3\gamma_0$$

There are four trilinear products $-i\gamma_0\gamma_1\gamma_2, -i\gamma_0\gamma_2\gamma_3, -i\gamma_0\gamma_3\gamma_1, -i\gamma_1\gamma_2\gamma_3$ and one quadrilinear product $\gamma_0\gamma_1\gamma_2\gamma_3 = \gamma_5$ which is called the pseudoscalar, it anticommutes with all the $\gamma_\mu$.

The gamma matrices are constructed as direct products of the familiar 2×2 Pauli matrices, which provide a spin-½ representation of the angular momentum algebra:

$$\sigma_1 = \begin{pmatrix} 0 & 1 \\ 1 & 0 \end{pmatrix}, \qquad \sigma_2 = \begin{pmatrix} 0 & -i \\ i & 0 \end{pmatrix}, \qquad \sigma_3 = \begin{pmatrix} 1 & 0 \\ 0 & -1 \end{pmatrix}, \qquad I = \sigma_0 = \begin{pmatrix} 1 & 0 \\ 0 & 1 \end{pmatrix}$$

The Pauli matrices anticommute and have a unit square:

$$\sigma_i\sigma_j = -\sigma_j\sigma_i, \qquad i \neq j$$

$$\sigma_i^2 = I$$

and also have the multiplication property $\sigma_i\sigma_j = i\sigma_k$, with $i,j,k$ in cyclic 1,2,3 order. Each gamma matrix is a direct product of Pauli matrices.

2-2-Representations of gamma matrices:

The standard or Pauli-Dirac or Bjorken-Drell representation:[3]

$$\gamma_0 = \begin{pmatrix} I_2 & 0 \\ 0 & -I_2 \end{pmatrix}, \qquad I_2 = \begin{pmatrix} 1 & 0 \\ 0 & 1 \end{pmatrix}$$

$$\gamma_i = \begin{pmatrix} 0 & \sigma_i \\ -\sigma_i & 0 \end{pmatrix} \quad j = 1,2,3$$

$$(\gamma_0)^2 = I, \qquad (\gamma_i)^2 = -I$$

Dirac matrices are $\beta = \begin{pmatrix} I_2 & 0 \\ 0 & -I_2 \end{pmatrix}$ and $\alpha_\iota = \begin{pmatrix} 0 & \sigma_i \\ \sigma_i & 0 \end{pmatrix}$ where $\gamma_0 = \beta$ and $\gamma_i = \beta\alpha_i$

Chiral or Weyl representation:

There are two choices for $\gamma_0$

$$\gamma_0 = \pm \begin{pmatrix} 0 & I_2 \\ I_2 & 0 \end{pmatrix}$$

$$\gamma_i = \begin{pmatrix} 0 & \sigma_i \\ -\sigma_i & 0 \end{pmatrix}$$

$$\gamma_5 = \pm \begin{pmatrix} 0 & I_2 \\ I_2 & 0 \end{pmatrix}$$

$$(\gamma_0)^2 = I, \qquad (\gamma_i)^2 = -I$$

Majorana representation:

In this representation, gamma matrices are imaginary, and the spinors are real.

$$\gamma_0 = \begin{pmatrix} 0 & \sigma_2 \\ \sigma_2 & 0 \end{pmatrix}, \qquad \gamma_1 = \begin{pmatrix} i\sigma_3 & 0 \\ 0 & i\sigma_3 \end{pmatrix}, \qquad \gamma_2 = \begin{pmatrix} 0 & -\sigma_2 \\ \sigma_2 & 0 \end{pmatrix}, \qquad \gamma_3 = \begin{pmatrix} -i\sigma_1 & 0 \\ 0 & -i\sigma_1 \end{pmatrix}$$

$$(\gamma_0)^2 = I, \qquad (\gamma_i)^2 = -I$$

Jauch-Rohrlich representation:

$$\gamma_0 = -i \begin{pmatrix} I_2 & 0 \\ 0 & -I_2 \end{pmatrix}, \qquad \gamma_i = \begin{pmatrix} 0 & \sigma_i \\ \sigma_i & 0 \end{pmatrix}$$

$$(\gamma_0)^2 = -I, \qquad (\gamma_i)^2 = I$$

$$\gamma_0 = -i\beta, \qquad \gamma_i = \alpha_i$$

### 2-3-Spacetime algebra:

The invariant interval of special relativity[5]

$$(ct)^2 - r^2 = (ct')^2 - (r')^2 = 0$$

is the fundamental algebraic concept we need to encode. The algebra we need to construct is generated by four orthogonal vectors $\{\gamma_0, \gamma_1, \gamma_2, \gamma_3\}$ satisfying the relations

$$\gamma_0^2 = 1, \qquad \gamma_0\gamma_i = 0, \qquad \gamma_i\gamma_j = -\delta_{ij}$$

where $i,j$=1,2,3. These are summarized in relativistic notation, using the 'particle physics' signature, as

$$\gamma_\mu\gamma_\nu = \eta_{\mu\nu} = \mathrm{diag}(+ - - -)$$

where $\mu,\nu$=0,1,2,3.

Throughout we use[1] Latin indices to denote the range 1 to 3 and Greek for the full spacetime range 0 to 3.

A general vector in the spacetime algebra can be constructed from the $\{\gamma_\mu\}$ vectors. A spacetime event, is encoded in the vector $x$, which has coordinates $x^\mu$ in the $\{\gamma_\mu\}$ frame. Explicitly, the vector $x$ is

$$x = x^\mu\gamma_\mu = ct\gamma_0 + x^i\gamma_i$$

We define the pseudoscalar $I$ by $I = \gamma_0\gamma_1\gamma_2\gamma_3$. As well as the four vectors, we also have four trivectors in our algebra. So, we arrive at an algebra with 16 terms. The $\{\gamma_\mu\}$ define an explicit basis for this algebra as follows: one scalar 1, four vectors $\{\gamma_\mu\}$, six bivectors $\{\gamma_\mu\gamma_\nu\}$, four trivectors $\{I\gamma_\mu\}$, and one pseudoscalar $I$. This is the spacetime algebra $\mathcal{G}(1,3)$. The structure of this algebra tells us practically all one needs to know about spacetime and the Lorentz transformation group. A general element of the spacetime algebra can be written as $M = \alpha + a + B + Ib + I\beta$, where $\alpha$ and $\beta$ are scalars, $a$ and $b$ are vectors and $B$ is a bivector.

These are the defining relations of the Dirac matrix algebra, except for the absence of an identity matrix on the right-hand side. It follows that the Dirac matrices define a representation of the spacetime algebra. In fact, the even subalgebra of the Dirac algebra is the spacetime algebra, and the Dirac algebra is the complexification of the spacetime algebra[8]

### 3- Spinors:

Spinors are introduced starting from the study of Clifford algebra. They arise naturally in discussions of the Lorentz group, as they are the most basic sort of mathematical objects that can be Lorentz transformed. Spinors as understood today are credited to Cartan (1913), they are closely related to Hamilton's quaternions[4]. Pauli formalized the connection between Cartan's work and the quantum physics of spin, where he modelled the spin using two-component complex vector and introducing Pauli spin matrices. Then for the required Lorentz covariant theory of spin, Dirac introduced a 4-component complex vector known as Dirac bispinor.[9]

As mentioned, the Clifford algebra associated with a vector space $V$ can be generated by $1, \gamma_1, \gamma_2, \gamma_3, \gamma_4$ satisfying the relation[10]

$$\gamma_a\gamma_b + \gamma_b\gamma_a = 2g_{ab} \quad (15)$$

and the elements of the algebra, understood as operators on the elements of a complex vector space $\mathfrak{D}$, which we will find in this section, are linear combinations with complex coefficients of $1, \gamma_1, \gamma_2, \gamma_3, \gamma_4$ and their products.

With a metric given of the following forms

$$(g_{ab}) = \begin{cases} \text{diag}(1,1,1,1): & \text{Euclidean signature} \\ \text{diag}(1,1,1,-1): & \text{Lorentzain signature} \\ \text{diag}(1,1,-1,-1): & \text{Kleininan signature} \end{cases}$$

equation (15) yields

$$(\gamma_a)^2 = \pm I_N \quad (16)$$

where $I_N$ denotes the identity of $\mathfrak{D}$, and

$$\gamma_a\gamma_b = -\gamma_b\gamma_a \quad \text{for } a \neq b \quad (17)$$

From equation (16), it follows that each operator $\gamma_a$ is invertible, $\gamma_a{}^{-1} = \pm\gamma_a$. And from equation (17), we obtain that $\det\gamma_a \det\gamma_b = (-1)^N \det\gamma_b \det\gamma_a$, where $N$ denotes the dimension of $\mathfrak{D}$. Hence, $N$ must be even.

---

[4] Quaternions are a system of extended complex numbers in four-dimensions, introduced by Hamilton (1845) to describe rotations in three-dimensional space.

Defining

$$\gamma_5 \equiv i^q \gamma_1 \gamma_2 \gamma_3 \gamma_4$$

where $q$ is the number of -1's appearing in $(g_{ab})$. We already learned that

$$\gamma_5 \gamma_a = -\gamma_a \gamma_5 \qquad (18)$$

and $\gamma_5{}^2 = (-1)^q \gamma_1{}^2 \gamma_2{}^2 \gamma_3{}^2 \gamma_4{}^2 = I_N$. Furthermore, from equation (18), $\gamma_5 = -\gamma_a \gamma_5 \gamma_a{}^{-1}$. Using the fact that tr($AB$)=tr($BA$), we obtain

$$\text{tr}(\gamma_5) = -\text{tr}(\gamma_a \gamma_5 \gamma_a{}^{-1}) = -\text{tr}(\gamma_5)$$

therefore

$$\text{tr}(\gamma_5) = 0 \quad (19)$$

. The eigenvalues of $\gamma_5$ equal to +1 or -1, and from equation (19), we conclude that the dimensions of $\mathfrak{D}$ is even. Hence, there is a basis of $\mathfrak{D}$ formed by eigenvectors of $\gamma_5$, with respect to which $\gamma_5$ is represented by

$$\begin{pmatrix} I_{N/2} & 0 \\ 0 & -I_{N/2} \end{pmatrix}$$

where $I_{N/2}$ is $N/2 \times N/2$ unit matrix. Writing $\gamma_a$ in block form

$$\begin{pmatrix} A_a & B_a \\ C_a & D_a \end{pmatrix}$$

where $A_a, B_a, C_a$ and $D_a$ are $N/2 \times N/2$ matrices. From equation (18), $A_a = D_a = 0$, and equation (15) amounts to

$$B_a C_b + B_b C_a = 2 g_{ab} I_{N/2}, \qquad C_a B_b + C_b B_a = 2 g_{ab} I_{N/2} \quad (20)$$

Equation (20) implies that $C_a = g_{aa} B_a{}^{-1}$. Furthermore, for $a \neq b$, $B_a C_b = -B_b C_a$, and therefore $\det B_a \det C_b = (-1)^N \det B_b \det C_a$. On the other hand, $\det C_a = (g_{aa})^{N/2} (\det B_a)^{-1}$

Assuming $N = 4$, let

$$B_a = \begin{pmatrix} x_a & y_a \\ z_a & w_a \end{pmatrix} \qquad (21)$$

where $a = 1,2,3,4$ and $x_a, y_a, z_a$ and $w_a$ are scalars, then

$$C_a = \lambda_a \begin{pmatrix} -w_a & y_a \\ z_a & -x_a \end{pmatrix} \qquad (22)$$

with $\lambda_a = -g_{aa} / \det B_a$.

Substituting equations (21) and (22) into (20), we found that all $\lambda_a$ must have the same value $\lambda$, and

$$\lambda(-x_a w_b - w_a x_b + y_a z_b + z_a y_b) = 2 g_{ab}$$

By applying the similarity transformation

$$\begin{pmatrix} 0 & B_a \\ C_a & 0 \end{pmatrix} \rightarrow \begin{pmatrix} \lambda^{1/2}I & 0 \\ 0 & I \end{pmatrix} \begin{pmatrix} 0 & B_a \\ C_a & 0 \end{pmatrix} \begin{pmatrix} \lambda^{-1/2}I & 0 \\ 0 & I \end{pmatrix} = \begin{pmatrix} 0 & \lambda^{1/2}B_a \\ \lambda^{-1/2}C_a & 0 \end{pmatrix}$$

and redefining $\lambda^{1/2}x_a, \lambda^{1/2}y_a, \lambda^{1/2}z_a, \lambda^{1/2}w_a$ as $x_a, y_a, z_a, w_a$ respectively, to get rid of $\lambda$. Therefore,

$$2g_{ab} = -x_a w_b - w_a x_b + y_a z_b + z_a y_b \quad (23)$$

$$\gamma_a = \begin{pmatrix} 0 & B_a \\ C_a & 0 \end{pmatrix} \qquad (24)$$

$$B_a = \begin{pmatrix} x_a & y_a \\ z_a & w_a \end{pmatrix}, \qquad C_a = \begin{pmatrix} -w_a & y_a \\ z_a & -x_a \end{pmatrix} \qquad (25)$$

Equation (23) implies that the (possibly complex) vectors with components $x_a, y_a, z_a, w_a$ form a null tetrad[5] that form a basis of (the complexification) of $V$. Any vector $v_a$ can be expressed as

$$v_a = g_{ab}v^b = \frac{1}{2}(-v^b w_b x_a - v^b x_b w_a + v^b z_b y_a + v^b y_b z_a)$$

Letting

$$\sigma_{a11'} \equiv x_a, \qquad \sigma_{a12'} \equiv y_a, \qquad \sigma_{a21'} \equiv z_a, \qquad \sigma_{a22'} \equiv w_a$$

equation (23) and (25) read[6]

$$\sigma_{aAB'}\sigma_b^{AB'} = -2g_{ab}$$

$$B_a = \begin{pmatrix} \sigma_{a11'} & \sigma_{a12'} \\ \sigma_{a21'} & \sigma_{a22'} \end{pmatrix}, \qquad C_a = -\begin{pmatrix} \sigma_a^{11'} & \sigma_a^{21'} \\ \sigma_a^{12'} & \sigma_a^{22'} \end{pmatrix}$$

The components of an element of $\mathfrak{D}$ with respect to a basis satisfying equations (23)-(25) is denoted by

$$\begin{pmatrix} \psi_A \\ \varphi^{A'} \end{pmatrix}$$

The elements of $\mathfrak{D}$ are called Dirac bispniors, and the gamma matrices operates on them as

$$\gamma_a \begin{pmatrix} \psi_A \\ \varphi^{A'} \end{pmatrix} = \begin{pmatrix} \sigma_{aAA'}\psi_A \\ \sigma_a^{AA'}\varphi^{A'} \end{pmatrix}$$

---

[5] cf. p. 23
[6] cf. p. 30

## III-Motivation to General Relativity

Consider an observer 'A' standing on the ground and watching a particle in free fall. This particle is only affected by the gravitational force of the Earth. This observer sees this particle falling in a straight line then 'A' will deduce that this particle is affected by some force driving it by a constant acceleration. Now if we consider a point-like observer 'B' fixed on the path of this particle. What 'B' observe is that the particle was accelerating towards him, then it reflected away from him. Then 'B' concludes that there is sharp potential reflecting this particle. Another example is that 'B' is moving parallel to the ground with constant velocity, he will see that particle moving in a parabolic path. This kind of 'fake' forces arises from coordinate transformations, and therefore can be eliminated by specific coordinate transformations. The 'real' gravitational force is associated with tidal forces that 'squeeze' the objects; as we will see eventually, squeezing space and time themselves.

1-Contravariant and Covariant Vectors:

The main variables we are dealing with in the context of general relativity are contravariant vectors and covariant vectors, each transform in a specific way. Considering two sets of coordinates $X^m(Y)$ and $Y^m(X)$ corresponding to two frames of reference $O$ and $O'$ respectively. A contravariant vector transforms in $O'$ as

$$V'^m = \sum_p \frac{\partial Y^m}{\partial X^p} V^p$$

Using the Einstein summation rule, indices appearing on one side of the equation are summed over, we write

$$V'^m = \frac{\partial Y^m}{\partial X^p} V^p$$

A covariant vector transforms as

$$W'_m = \frac{\partial X^p}{\partial Y^m} W_p$$

To give an insight to the transformation of contravariant vectors, such as velocity, the components of the vector in $O'$ ($V'^m$) equals to the variations of the bases of $O$ ($dX^p$)with respect to those of $O'$ ($Y^m$) times the corresponding component of $V^m$ in $O$ ($V^p$). For the covariant vectors, considering the gradient $\frac{\partial S}{\partial X^m}$ in $O$. In $O'$ we get

$$\frac{\partial S}{\partial Y^m} = \frac{\partial S}{\partial X^p}\frac{\partial X^p}{\partial Y^m} = \frac{\partial X^p}{\partial Y^m}\left(\frac{\partial S}{\partial X^p}\right)$$

Concerning tensors, if we consider a rank 2 contravariant tensor defined by $V^m U^n = T^{mn}$, it transforms as

$$T'^{mn} = V'^m U'^n = \frac{\partial Y^m}{\partial X^p}\frac{\partial Y^n}{\partial X^q} V^p U^q$$

In general relativity, geometry is not necessarily flat, so the distance between two neighboring point in a curved geometry is $ds^2 = g_{mn}(x)dx^m dx^n$, where $g_{mn}$ is called the metric.

On the difference between contravariant and covariant vectors, considering the case of a vector $V$ written in the conventional way $V = V^1 e_1 + V^2 e_2 + \cdots = V^m e_m$. $e_m$ are basis vectors, while $V^m$ are the contravariant component of $V$. If we project $V$ on $e_p$ in flat coordinates we get $V^p$, but in general, $V_p = V.e_p = V^m e_m.e_p = V^m g_{mp}$. The length of $V$ is

$$V.V = (V^m e_m).(V^n e_n) = V^m V^n e_m.e_n = V^m V^n g_{mn}$$

We conclude that the metric $g_{mn}$ is related to the definition and computation of lengths, and the vector is constructed from the contravariant components, while the covariant components are the dot product with the basis vectors; for which the covariant and contravariant components are the same in Cartesian coordinates.

<u>2-Tensors, Metrics, Covariant Derivatives:</u>

Tensor equation with balanced indices, e.g., $T_a^b = U_a^b$, hold true in all frames. That is the significance of using tensors in general relativity, where all the physical laws most hold true regardless of the observer.

Tensors can be added as $T_{...p}^{m...} \pm S_{...p}^{m...} = (T \pm S)_{...p}^{m...}$ and multiplied as $V^m \otimes W_n = T_n^m$, later we will consider differentiation of tensors.

The infinitesimal displacement in $O$ and $O'$ is respectively $ds^2 = g_{mn}(x)dx^m dx^n$ and $ds^2 = g'_{pq}(y)dy^p dy^q$. If

$$dx^m = \frac{\partial x^m}{\partial y^p}dy^p$$

then

$$ds^2 = g_{mn}(x)\frac{\partial x^m}{\partial y^p}\frac{\partial x^n}{\partial y^q}dy^p dy^q$$

So, the metric tensor transforms as

$$g'_{pq}(y) = \frac{\partial x^m}{\partial y^p}\frac{\partial x^n}{\partial y^q}g_{mn}(x)$$

$g_{mn}g^{pn} = \delta_m^p$, where $\delta_m^p$ is the identity matrix , or Kronecker delta, therefore $g^{pn}$ is the inverse of $g_{mn}$. The metric tensor is used for lowering or rising indices: $V^m g_{mn} = V_n$ and $V_n g^{mn} = V^m$.

If the metric tensor can be transformed by coordinates transformation to be the Kronecker delta, i.e., the identity matrix, then this geometry is flat. That is equivalent, as we will see, to a zero Riemann curvature tensor.

Considering an arbitrary curved coordinate system. We define Gaussian normal coordinates at a point $x_0$. At $x = x_0$ we have

$$g_{mn} = \delta_{mn}, \qquad \frac{\partial g_{mn}}{\partial x^r} = 0, \qquad \frac{\partial^2 g_{mn}}{\partial x^r \partial x^s} = 0$$

Considering a vector field $V^m$ in these coordinate systems. In Gaussian normal coordinates $\frac{\partial V^m}{\partial x^r} = 0$, while in the curved coordinates it is not zero, because the coordinate itself is varying. Therefore, we are not dealing with a tensor, since if a tensor equation is zero in a coordinate system, then it is zero in every coordinate system. So, we define a new way of differentiation, called the covariant derivative defined by

$$D_r V_m = \partial_r V_m - \Gamma_{rm}^t V_t \quad (26)$$

This derivative is equivalent to differentiating a vector in normal Gaussian coordinates then transforming into another coordinate system. The covariant derivative $D_r$ is a tensor.

The first term '$\partial_r V_m$' represents the change of the vector field, and the second term '$\Gamma_{rm}^t V_t$' represents the change in coordinates. $\Gamma_{rm}^t$ is called connection coefficient or Christoffel symbol, which is not a tensor, defined by

$$\Gamma_{rm}^t = \frac{1}{2} g^{st} (\partial_n g_{mn} + \partial_m g_{mn} - \partial_s g_{mn})$$

For contravariant vectors,

$$D_m V^n = (D_m g^{np}) V^p + g^{np} (D_m V_p) = g^{np} (D_m V_p)$$

In flat space $D_r D_s V_n = D_s D_r V_n$, equivalently,

$$D_r D_s V_n - D_s D_r V_n = 0 \quad (27)$$

$$\Rightarrow D_r D_s V_n = D_s (\partial_r V_n - \Gamma_{rn}^t V_t)$$

We aim to define a tensor for which if it is zero, then the space is flat. Let

$$D_r D_s V_n - D_s D_r V_n = R_{srn}^t V_t$$

By equations (26) and (27), we get the Riemann curvature tensor

$$R_{srn}^t = \partial_r \Gamma_{sn}^t - \partial_s \Gamma_{rn}^t + \Gamma_{sn}^p \Gamma_{pr}^t - \Gamma_{rn}^p \Gamma_{ps}^t$$

Riemann tensor $R_{ijkl}$ is antisymmetric in both pairs of indices ($ij$) and ($kl$) and it is unchanged by simultaneous interchange of the two pairs. By virtue of this symmetry, the only nontrivial contraction we can make, by raising and index and contracting it with one of the covariant indices, is that leading to Ricci tensor. Accordingly, it will be convenient to separate Riemann tensor into a 'trace-free' part and a 'Ricci' part. This separation is accomplished by Weyl tensor[11]

$$C_{ijkl} = R_{ijkl} - \frac{1}{n-2} \big( g_{ik} R_{jl} + g_{il} R_{ik} - g_{jk} R_{il} - g_{il} R_{jk} \big) + \frac{1}{(n-1)(n-2)} \big( g_{ik} g_{jl} - g_{il} g_{jk} \big) R$$

This tensor has all the symmetry properties of Riemann tensor, but $g^{jl}C_{ijkl} = 0$ in contrast to $g^{jl}R_{ijkl} = R_{ik}$.

We define Ricci identity by[11]

$$R_{jkl}^{i}Z_i = Z_{j;k;l} - Z_{j;l;k}$$

Where the semicolon notation represents the covariant derivative with respect to this index, e.g., $Z_{i;j} \equiv D_j Z_{ij}$.

### 3-Parallel transport and geodesics:

Considering a vector field $V^n$, if it is parallel to itself then its covariant derivative is zero.

$$DV^n = D_m V^n dx^m = \frac{\partial V^n}{\partial x^m} dx^m + \Gamma_{mr}^n V^r dx^m$$

$$\Rightarrow DV^n = \partial V^n + \Gamma_{mr}^n V^r dx^m$$

If a vector is parallel transported on a particular path, then $DV^n = 0$.

The shortest path between two points is called a geodesic. A more formal definition is that the tangent vector along a geodesic is constant. The tangent vector is defined by

$$t^m = \frac{dx^m}{ds} \qquad (28)$$

with $ds^2 = g_{rs} dx^r dx^s$. The notion of geodesic is

$$dt^n + \Gamma_{mr}^n t^r dx^m = 0 \qquad (29)$$

Dividing equation (4) by $ds$

$$\frac{dt^n}{ds} = -\Gamma_{mr}^n t^r \frac{dx^m}{ds} = -\Gamma_{mr}^n t^r t^m \qquad (30)$$

Equation (30) is the equation of motion. Substituting equation (28) in equation (30), we get

$$\frac{d^2 x^n}{ds^2} = -\Gamma_{mr}^n t^r t^m$$

This is Newton's second law in a gravitational field. Note that $\Gamma_{mr}^n$ contains the metric tensor, which is related to gravity. So, a particle in a gravitational field moves in the shortest path in the spacetime.

### 4- Spacetime geometry and Schwarzschild black hole:

The geometry of general relativity is pseudo-Riemannian geometry, meaning that the distance between two neighboring points could be negative. In special relativity, the spacetime is flat, its geometry is called Minkowskian, with metric tensor

$$g_{\mu\nu} = \begin{pmatrix} -1 & 0 & 0 & 0 \\ 0 & 1 & 0 & 0 \\ 0 & 0 & 1 & 0 \\ 0 & 0 & 0 & 1 \end{pmatrix}$$

The shortest distance between two points is called proper time,

$$d\tau^2 = dt^2 - dx^2 - dy^2 - dz^2 = -ds^2 = -g_{\mu\nu}dx^\mu dx^\nu$$

With $x^\mu = (t, x, y, z) \equiv (x^0, x^1, x^2, x^3)$. The metric of general relativity is spacetime dependent $g_{\mu\nu} = g_{\mu\nu}(x)$. The fact that the metric has one negative eigenvalue and 3 positive eigenvalues or vice versa, means that there is one dimension of time and three of spaces. Metrics of the following form are not allowed.

$$\begin{pmatrix} -1 & 0 & 0 & 0 \\ 0 & -1 & 0 & 0 \\ 0 & 0 & 1 & 0 \\ 0 & 0 & 0 & 1 \end{pmatrix}$$

If $d\tau^2$ is positive, we say it is timelike path, if negative it is spacelike path, and if null it is lightlike path.

Recall the polar and hyperbolic coordinates,

$$x = r\cos\theta, \qquad y = r\sin\theta, \ x^2 + y^2 = r^2$$

$$\cos\theta = \frac{e^{i\theta} + e^{-i\theta}}{2}, \qquad \sin\theta = \frac{e^{i\theta} - e^{-i\theta}}{2i}$$

With $ds^2 = r^2 d\theta^2 + dr^2$.

$$x = r\cosh\omega, \qquad t = r\cosh\omega, \qquad x^2 - t^2 = r^2$$

$$\cosh\omega = \frac{e^\omega + e^{-\omega}}{2}, \qquad \sinh\omega = \frac{e^\omega - e^{-\omega}}{2}$$

Where $\omega = ]-\infty, +\infty[$ is the hyperbolic angle, with $d\tau^2 = r^2 d\omega^2 - dr^2$.

In polar coordinates, it is obvious that a path along a fixed $r$ is equivalent to uniform acceleration. By analogy, the fixed $r$ is hyperbolic coordinates is equivalent to a path under constant acceleration. The acceleration felt along a fixed $r$ is called the proper acceleration. For paths of small $r$, the acceleration is large.

Schwarzschild metric is defined by

$$d\tau^2 = \left(1 - \frac{R_s}{r'}\right)^2 dt^2 - \frac{dr'^2}{\left(1 - \frac{R_s}{r'}\right)} - r'^2 d\Omega^2$$

Where, $d\Omega^2 = d\theta^2 + cos^2\theta \ d\varphi^2$ is the metric on a sphere, and $R_s = 2MG$ is the Schwarzschild radius of a black hole of mass $M$.

Let $r = \frac{r\prime}{R_s}$ and $t = \frac{t\prime}{R_s}$, then

$$d\tau^2 = \left(\frac{r-1}{r}\,dt^2 - \frac{r}{r-1}\,dr^2 - r^2 d\Omega^2\right)R_s$$

For a unit black hole, $R_s = 1$. Considering the radial path only, that is at the same time $dt$=0 and $d\omega$=0, we get $ds = dr\sqrt{\frac{r}{r-1}}$. Let $\rho(r)$ be the proper distance from the horizon,

$$\rho = \int_1^r dr\sqrt{\frac{r}{r-1}} \Rightarrow \frac{d\rho^2}{d^2r} = \frac{r}{r-1}$$

If we let $t = 2\omega \Longrightarrow dt = 2d\omega \Longrightarrow dt^2 = 4d\omega^2$, then

$$d\tau^2 = F(\rho)\rho^2 d\omega^2 - d\rho^2 - r^2(\rho)d\Omega^2$$

The horizon corresponds to $\rho$=0.

$$\lim_{\rho\to\infty} F(\rho)\rho^2 = 4, \qquad \lim_{\rho\to 0} F(\rho) = 1, \qquad \lim_{\rho\to 0} r(\rho) = 1 + \frac{\rho^2}{4}$$

For small $\rho$, $d\tau^2 = \rho^2 d\omega^2 - d\rho^2$

The following picture represents Kruskal coordinates. Region I represents our universe. The line of $\omega = \infty$ represents the event horizon of Schwarzschild black hole. Beyond this line, the spacetime is space-like. Region II is the interior of the black hole. The $r = 0$ line represents the singularity, notice that the singularity is a moment of time, rather that a point in space. Region III has no physical meaning but can be interpreted as a 'parallel universe', i.e., it is an inaccessible region with the same laws of physics than region I. Region IV is a parallel image of region II. Though it has no physical meaning, but can be interpreted as a 'white hole'

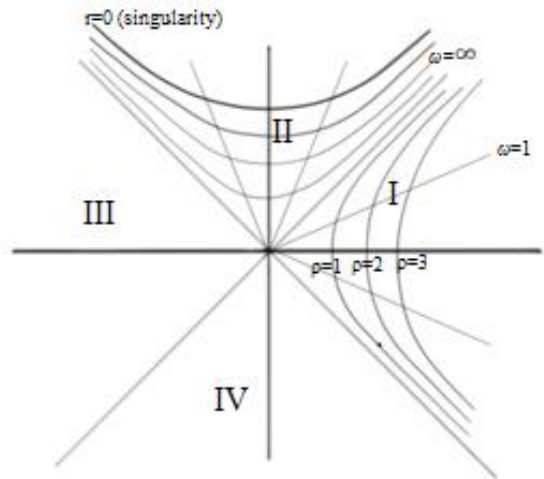

<u>5-Einstein Field Equation:</u>

In this last subsection, Einstein field equations are derived from the Newtonian picture, an approach known as 'primitive' field theoretic approach.

The gravitational force of a particle of mass $m$ in a gravitational potential $\varphi$ is

$$F = -m\nabla\varphi$$

As $F = ma$, then $a = -\nabla\varphi$: this can be understood as the gravitational potential tells the particle how to move. On the other hand, Poisson's equation

$$\nabla^2\varphi = 4\pi G\rho$$

can be understood as how a gravitational source of density $\rho$ determines the gravitational field. This interrelation is essential to understand the idea behind Einstein's field equations.

In analogy with electricity, the charge density and the current are conserved by a continuity equation $\nabla.J + \frac{\partial\sigma}{\partial t} = 0$, where $J$ is the current and $\sigma$ is the charge density. These variables can be put together by a four-vector $J^\mu = (J^0 = \sigma, J^1 = J^x, J^2, J^3)$. So, we get $\frac{dJ^\mu}{\partial x^\mu} = 0$.

Energy and momentum are conserved and can flow, we construct the four-vector

$p^\mu = (p^0 = E, p^m)$ where $m$=1,2,3. To understand the flows of energy and momentum and to construct the equations in a way to be true in any frame of reference, we introduce the energy-momentum tensor $T^{\mu\nu}$. The density of energy is described by $T^{00}$, the flow of energy in the $x$ direction is described by $T^{01}$, the density of the $m^{\text{th}}$ momentum is $T^{m0}$ and its flow in the $n^{\text{th}}$ direction is $T^{mn}$. The energy-momentum or the stress-energy tensor $T^{\mu\nu}$ is symmetric, i.e., $T^{\mu\nu} = T^{\nu\mu}$, which means that the flow of energy in a direction is equivalent to the density of momentum in this direction. The appropriate continuity equation is

$$D_\mu T^{\mu\nu} = 0$$

The component $g^{00}$ of the metric $g^{\mu\nu}$ outside a massive object is

$$g^{00} = 1 - \frac{2MG}{r} = 1 - 2\varphi \Rightarrow \overbrace{\nabla^2 g^{00}}^{geometry} = \overbrace{8\pi G\rho}^{matter}$$

So, the Einstein field equation must be of the form $G^{\mu\nu} = 8\pi G T^{\mu\nu}$, with $G^{\mu\nu}$ involving a second derivative of the metric. Recall Riemann curvature tensor

$$R^\sigma_{\mu\nu\tau} = \partial_\mu\Gamma^\sigma_{\nu\tau} - \partial_\nu\Gamma^\sigma_{\mu\tau} + \Gamma^\sigma_{\nu\lambda}\Gamma^\lambda_{\mu\tau} - \Gamma^\sigma_{\mu\lambda}\Gamma^\lambda_{\nu\tau}$$

With the Christoffel symbol

$$\Gamma^\sigma_{\nu\tau} = \frac{1}{2}g^{\sigma\delta}(\partial_\tau g_{\delta\nu} + \partial_\nu g_{\delta\tau} - \partial_\delta g_{\nu\tau})$$

The contraction of Riemann tensor is the Ricci tensor $R_{\mu\nu}$. The curvature scalar $R = R^\mu_\mu = R^{\mu\tau}g_{\mu\tau}$.

We have two choices that make sense: $R^{\mu\nu} = 8\pi G T^{\mu\nu}$ or $g^{\mu\nu} R = 8\pi G T^{\mu\nu}$

From the continuity equation $D_\mu T^{\mu\nu} = 0 \longrightarrow D_\mu G^{\mu\nu} = 0$.

$$D_\mu(g^{\mu\nu} R) = \underbrace{(D_\mu g^{\mu\nu})}_{0} R + g^{\mu\nu}(D_\mu R) = g^{\mu\nu}(\partial_\mu R) \neq 0$$

$$D_\mu R^{\mu\nu} = \frac{1}{2} g^{\mu\nu} \partial_\mu R$$

So, $R^{\mu\nu} - \frac{1}{2} g^{\mu\nu} R = G^{\mu\nu}$, that is Einstein tensor.

Therefore, the Einstein field equation reads

$$R^{\mu\nu} - \frac{1}{2} g^{\mu\nu} R = 8\pi G T^{\mu\nu}$$

**III- Dirac equation in Newman-Penrose Formalism:**

In this section we develop the mathematical formalism in which the Dirac equation is imposed. Firstly, the tetrad formalism of general relativity is introduced, then a special yet befitting formalism is adopted, the Newman-Penrose formalism. Lastly, a spinorial basis for Newman-Penrose formalism is found to fit with Dirac spinor analysis.

1-Tetrad Formalism:

1-1-Tetrad Field:

The standard way of treating problems in general relativity used to be to consider Einstein field equation in a local coordinate basis adapted to the problem. By considering basis vectors that are not derived from any coordinate system, the relationship of variables in general relativity, such as connection and curvature, with gauge theories of particle physics can be more transparent[11]. So, by projecting the relevant quantities on the chosen basis, one can consider the equations satisfied by them.

The basis vector is chosen to be orthogonal, in a sense that is appropriate to the signature of the spacetime. That is, if the canonical form of the metric is $\eta_{AB}$, the inner product of the basis vector is required to be

$$g(\boldsymbol{e}_A, \boldsymbol{e}_B) = \eta_{AB} \quad (31)$$

where $g(\,,)$ is the usual metric tensor. The orthogonal (non-coordinate) basis $\boldsymbol{e}_A$ is related to the old coordinate basis $\boldsymbol{e}_\alpha$ by

$$\boldsymbol{e}_\alpha = e_\alpha^A \boldsymbol{e}_A$$

The components $e_\alpha^A$ from an $n \times n$ invertible matrix, an referred to as the tetrad[7] or veilbein[8]. We denote their inverse $e_A^\alpha$, which satisfy

$$e_\alpha^A e_A^\beta = \delta_\alpha^\beta, \quad e_\alpha^A e_B^\alpha = \delta_B^A \quad (32)$$

The Greek indices are tetrad indices and the Lattin indices are tensor indices.

The components of $\boldsymbol{e}_A$ in the coordinate basis is $\boldsymbol{e}_A = e_A^\alpha \boldsymbol{e}_\alpha$

Equation (31) becomes

$$g_{\alpha\beta} e_B^\alpha e_A^\beta = \eta_{AB} \quad (33)$$

This equation leads to think about tetrad as 'square root' of the metric[12]. Tetrads can be interpreted as a set of four linearly independent vectors defined at each point in the semi-Riemannian spacetime. The choice of tetrad field determines the metric, by equation (33)

---

[7] From the Greek *tetras,* "a group of four"
[8] The German for "many legs"

$$g_{\alpha\beta} = e_\alpha^A e_\beta^B \eta_{AB}, \quad \eta_{AB} = e_A^\alpha e_B^\beta g_{\alpha\beta}$$

To obtain the tetrad components of any tensor field, we project it into the tetrad frame

$$A_\alpha = e_{\alpha A} A^A$$

$$A^\alpha = \eta^{\alpha\beta} A_\beta = e_A^\alpha A^A = e^{\alpha A} A_A$$

$$A^A = e_\alpha^A A^\alpha = e^{\alpha A} A_\alpha$$

More generally,

$$T_{\alpha\beta} = e_\alpha^A e_\beta^B T_{AB}$$

### 1-2-Spinor Affine Connection and Spin Connection:

For writing Dirac equation in general relativity, we need to introduce spacetime-dependent matrices $\bar\gamma^\alpha(x)$ defined by $\bar\gamma^\alpha(x) = e_A^\alpha \gamma^A$.

Since $\{\gamma^A, \gamma^B\} = 2\varepsilon\eta^{AB}$, therefore

$$\{\bar\gamma^A, \bar\gamma^B\} = 2\varepsilon g^{AB} \quad (34)$$

where $\varepsilon = \pm 1$. The matrices anticommute for $\alpha \neq \beta$ only if the metric is diagonal.

In this context, we define the spinor $\psi$ as a quantity that transforms as

$$\tilde\psi_e = L\psi_h$$

The derivative of a spinor is not a spinor since

$$\tilde\psi_{,\mu} = L\psi_{,\mu} + L_{,\mu}\psi$$

where the comma notation means that the quantity is differentiated with respect to $x^\mu$, i.e., $V_{,\mu} \equiv \frac{\partial V}{\partial x^\mu}$. So, we define the covariant derivative of spinor by

$$D_\mu\psi = \psi_{,\mu} + \Gamma_\mu\psi$$

with the spinor affine connection $\Gamma_\mu$ to be determined. Under the assumption that the operator $D_\mu$ is a derivative, it may be extended as an operator on a matrix-valued field $M$. By writing $M$ as a linear combination of tensor products of covectors with vectors, then we have that

$$D_\mu M = \partial_\mu M + [\Gamma_\mu, M] \quad (35)$$

We impose the additional requirement that $D_\mu$ is metric compatible, i.e.,

$$D_\mu g^{\alpha\beta} = 0 \quad (36)$$

By equation (34), equation (36) is equivalent to $D_\mu(\{\bar\gamma^\alpha, \bar\gamma^\beta\}) = 0$ and a sufficient condition follows

$$D_\mu \bar{\gamma}^\nu = 0 \quad (37)$$

Using the definition of the covariant derivative and equations (35) and (36),

$$D_\mu \bar{\gamma}^\nu = \partial_\mu \bar{\gamma}^\nu + [\Gamma_\mu, \bar{\gamma}^\nu] = \bar{\gamma}^\nu_{,\mu} + \Gamma^\nu_{\lambda\mu} \bar{\gamma}^\lambda + \Gamma_\mu \bar{\gamma}^\nu - \bar{\gamma}^\nu \Gamma_\mu = 0$$

To differential spinors, we introduce the spin connection $\omega^A_{\mu B}$ defined by[12]

$$\nabla_\mu X^A_B = \partial_\mu X^A_B + \omega^A_{\mu C} X^C_B - \omega^C_{\mu B} X^A_C$$

One can derive a relation between the spin connection, the tetrads, and the spinor connection $\Gamma^\nu_{\lambda\mu}$, for which no assumptions are needed,

$$\Gamma^\nu_{\lambda\mu} = e^\nu_A \, \partial_\mu e^A_\lambda + e^\nu_A e^B_\lambda \, \omega^A_{\mu B}$$

or equivalently[3],

$$\omega^A_{\mu B} = -e^\nu_B \big( \partial_\mu e^A_\lambda - \Gamma^\lambda_{\mu\nu} e^A_\lambda \big)$$

It follows that,

$$\omega_{AB\mu} = e_{A\beta} \nabla_\mu e^\beta_B = g_{\beta\alpha} e^\alpha_A \nabla_\mu e^\beta_B = \eta_{AC} e^C_\beta \nabla_\mu e^\beta_B$$

$\omega_{AB\mu}$ are antisymmetric in $A$ and $B$: $\omega_{AB\mu} = -\omega_{BA\mu}$, and $\omega_{AB\mu} = \omega_{\mu AB}$.[9]

$\Gamma_\mu$ satisfies the relation

$$\Gamma_\mu = \frac{\varepsilon}{4} \omega_{AB\mu} \gamma^A \gamma^B = \frac{\varepsilon}{2} \omega_{AB\mu} \Sigma^{AB}$$

where, $\Sigma^{AB} = \frac{1}{2} [\gamma^A, \gamma^B]$.

### 1-3-Directional Derivatives and Ricci Rotation Coefficients:

The contravariant vectors $e_A$, considered as tangent vectors, define the directional derivatives[11]

$$e_A = e^\mu_A \frac{\partial}{\partial x^\mu}$$

Then,

$$\varphi_{,A} = e^\mu_A \frac{\partial \varphi}{\partial x^\mu} = e^\mu_A \varphi_{,\mu}$$

where $\varphi$ is a scalar field. More generally,

$$A_{A,B} = e^\mu_B \frac{\partial}{\partial x^\mu} A_A = e^\mu_B \frac{\partial}{\partial x^\mu} e^\nu_A A_\nu \equiv e^\mu_B \nabla_\mu (e^\nu_A A_\nu) = e^\mu_B \big( e^\nu_A A_{\nu;\mu} + A_\lambda e^\lambda_{A;\mu} \big)$$

---

[9] Note that the metric signature affects the sign of $\omega_{AB\mu}$.

$$\Rightarrow A_{A,B} = e_A^\nu A_{\nu;\mu} e_B^\mu + e_{A\lambda;\mu} e_B^\mu e_C^\mu A^C \quad (38)$$

With the definition

$$\gamma_{ABC} := e_{A\lambda;\mu} e_B^\mu e_C^\mu$$

equation (38) can be written as

$$A_{A,B} = e_A^\nu A_{\nu;\mu} e_B^\mu + \gamma_{ABC} A^C \quad (39)$$

where $\gamma_{ABC}$ is the Ricci rotation coefficients. An equivalent definition is

$$e_{A\lambda;\mu} = e_\lambda^C \gamma_{ABC} e_\mu^B$$

Ricci rotation coefficients are related to the spin connection by[3]

$$\gamma_{ABC} = \omega_{AB\mu} e_C^\mu$$

Ricci rotation coefficients are antisymmetric in the first pair on indices[11], $\gamma_{ABC} = -\gamma_{BAC}$, which follows from expending the identity:

$$0 = \eta_{AB;\mu} = [e_{A\nu} e_B^\nu]_{;\mu}$$

note that $e_{A;\mu}^\nu = -\gamma_{A\mu}^\nu$

Therefore, equation (39) can be written as

$$e_A^\mu A_{\mu,\nu} e_B^\nu = A_{A,B} - \eta^{CD} \gamma_{CAB} A_D \quad (40)$$

The quantity on the RHS is called the intrinsic derivative of $A_A$ in the direction of $e_B$, and written as

$$A_{A|B} = e_A^\mu A_{\mu,\nu} e_B^\nu$$

Projecting Ricci identity,

$$R_{\mu\nu\sigma\tau} e_A^\mu = e_{A\nu;\sigma;\tau} - e_{A\nu;\tau;\sigma}$$

into the tetrad frame, we get

$$R_{ABCD} = R_{\mu\nu\sigma\tau} e_A^\mu e_B^\nu e_C^\sigma e_D^\tau$$

$$= \left\{ -[\gamma_{AEF} e_\nu^E e_\sigma^F]_{;\tau} + [\gamma_{AEF} e_\mu^E e_\tau^F]_{,\sigma} \right\} e_B^\mu e_C^\sigma e_D^\tau$$

Now we get Riemann tensor in terms of Ricci rotation coefficients,

$$R_{ABCD} = \gamma_{ABC,D} + \gamma_{ABD,C} + \gamma_{BAE}[\gamma_{CD}^E - \gamma_{DC}^E] + \gamma_{EAC}\gamma_{BD}^E - \gamma_{EAD}\gamma_{BC}^E$$

We define the Fock-Ivenanko coefficients $\Gamma_C$ in terms of the spinor affine connection $\Gamma_\mu$ by[3]

$$\Gamma_C = e_C^\mu \Gamma_\mu$$

and in terms of Ricci rotation coefficients by

$$\Gamma_C = \frac{\varepsilon}{4} \Gamma_{ABC} \gamma^A \gamma^B$$

Now we can define the covariant derivative $D_C$ to be used in Dirac equation by

$$D_C \psi = e_C \psi + \Gamma_C \psi$$

For a spin-½ particle of mass $m$, Dirac equation in curved spacetime is

$$i\gamma_C D_C \psi - m\psi = 0 \quad (41)$$

Which is equivalent to

$$\bar{\gamma}^\mu (\partial_\mu + \Gamma_\mu)\psi - m\psi = 0 \Rightarrow \bar{\gamma}^\mu D_\mu \psi - m\psi = 0$$

## 2-Newman-Penrose Formalism:

The Newman-Penrose formalism is a tetrad formalism with the special choice of basis vectors to be null and consisting of a pair of real vectors $l$ and $n$, and a pair of complex conjugate vectors $m$ and $\bar{m}$. The underlying motivation of such a choice was Penrose belief that the essential element of the spacetime is the light cone structure, which makes possible the introduction of spinorial analysis. It will appear that the light cone structure of the spacetime of black hole solutions is exactly of the kind that makes the Newman-Penrose formalism most effective for grasping the inherent symmetries and the analytical richness[11].

Starting by a set of orthonormal tetrad vector field $e_A$ satisfying[3]

$$\eta_{AB} = e_A^\alpha e_B^\beta g_{\alpha\beta}$$

Considering the basis $\{\lambda_1, \lambda_2, \lambda_3, \lambda_4\}$ with respect to the metric tensor of a vector field $V$ of the from[4]

$$\begin{pmatrix} 0 & 1 & 0 & 0 \\ 1 & 0 & 0 & 0 \\ 0 & 0 & 0 & 1 \\ 0 & 0 & 1 & 0 \end{pmatrix}$$

where the only non-vanishing inner products among the vectors $\lambda_\mu$ are given by

$$g(\lambda_1, \lambda_2) = 1 = g(\lambda_3, \lambda_4)$$

Such basis is null since $g(\lambda_\mu, \lambda_\mu) = 0 \;\; \forall \mu$. Only when the metric is Kleinian, i.e., $(g_{\mu\nu}) = (1,1,-1,-1)$, it will be possible to find real null tetrad, so we assume that $\lambda_\mu$ belong to the complexification of $V$.

We define the complex null tetrad $l, n, m, \bar{m}$ by[3]

$$\lambda_1 = l = \frac{1}{\sqrt{2}} (e_0 + e_3), \qquad \lambda_2 = n = l = \frac{1}{\sqrt{2}} (e_0 - e_3)$$

$$\lambda_3 = m = \frac{1}{\sqrt{2}}(e_1 + ie_2), \qquad \lambda_3 = \bar{m} = \frac{1}{\sqrt{2}}(e_1 - ie_2)$$

satisfying the orthogonality condition

$$l^\mu m_\mu = n^\mu m_\mu = l^\mu \bar{m}_\mu = n^\mu \bar{m}_\mu = 0$$

and

$$l^\mu l_\mu = n^\mu n_\mu = m^\mu m_\mu = \bar{m}^\mu \bar{m}_\mu = 0$$

and the normalization condition

$$l^\mu n_\mu = 1, \qquad m^\mu \bar{m}_\mu = -1$$

This set of null tetrad transform under three classes of transformations. A null rotation:

$$l' = l, \qquad m' = m + Bl, \qquad n' = n + B^*m + B\bar{m} + BB^*l$$

another null rotation,

$$n' = n, \qquad m' = m + Bl, \qquad l' = l + B^*m + B\bar{m} + BB^*n$$

and boost and orthogonal rotation,

$$l' = Al, \qquad m' = e^{i\varphi}m, \qquad n' = A^{-1}n$$

where $A$ and $\varphi$ are real functions, $B$ is a complex function and $B^*$ is its complex conjugate functions.

The null tetrad vectors are designated as directional derivatives

$$l = D, \qquad n = \Delta, \qquad m = \delta, \qquad \bar{m} = \delta^*$$

The role of covariant derivative $D_\alpha$ is taken over in the Newman-Penrose formalism by four scalar operators

$$D = l^\alpha D_\alpha, \qquad \Delta = n^\alpha D_\alpha, \qquad \delta = m^\alpha D_\alpha, \qquad \bar{\delta} = \bar{m}^\alpha D_\alpha$$

Ricci rotation coefficients are called spin coefficients in Newman-Penrose formalism[11], denoted by

$$\kappa = \gamma_{311}, \qquad \rho = \gamma_{314}, \qquad \varepsilon = \frac{1}{2}(\gamma_{211} + \gamma_{341}), \qquad \sigma = \gamma_{313}, \qquad \mu = \gamma_{243},$$

$$\gamma = \frac{1}{2}(\gamma_{212} + \gamma_{342}), \qquad \lambda = \gamma_{244}, \qquad \tau = \gamma_{312}, \qquad \alpha = \frac{1}{2}(\gamma_{214} + \gamma_{344}),$$

$$\beta = \frac{1}{2}(\gamma_{213} + \gamma_{343}), \qquad \nu = \gamma_{242}, \qquad \pi = \gamma_{241}$$

The functions $A$, $B$, and $\varphi$ are chosen in a way to maximize the number of vanishing spin coefficients.

## 3-Dirac Equation in Newman-Penrose Formalism:

In this section, the spinorial basis of Newman-Penrose formalism is elaborated, in which the form of Dirac equation is suitable for calculations.

### 3-1-Dyad Field:

The notion of spinors originates in the observation that a 4-vector in Minkowski space can be represented equally by a Hermitian matrix, and that a unimodular transformation in the complex two-dimensional space induces a Lorentz transformation in Minkowski space.[11]

Consider in Minkowski space the vector $x^\mu$ and let

$$(x^0)^2 - (x^1)^2 - (x^2)^2 - (x^3)^2 = 0 \quad (42)$$

$x^\mu$ can be represented in terms of two complex numbers $\xi^0$ and $\xi^1$, and their complex conjugates $\bar\xi^{0'}$ and $\bar\xi^{1'}$ as

$$x^0 = \frac{1}{\sqrt 2}\left(\xi^0\bar\xi^{0'} + \xi^1\bar\xi^{1'}\right), \qquad x^1 = \frac{1}{\sqrt 2}\left(\xi^0\bar\xi^{1'} + \xi^1\bar\xi^{0'}\right), \qquad x^2 = -\frac{1}{\sqrt 2}\left(\xi^0\bar\xi^1 - \xi^1\bar\xi^{0'}\right) \left.\vphantom{\begin{matrix}a\\b\end{matrix}}\right\}$$

$$x^3 = \frac{1}{\sqrt 2}\left(\xi^0\bar\xi^{0'} - \xi^1\bar\xi^{1'}\right) \qquad (43)$$

or equivalently,

$$\xi^0\bar\xi^{0'} = \frac{1}{\sqrt 2}(x^0 + x^3), \qquad \xi^0\bar\xi^{1'} = \frac{1}{\sqrt 2}(x^1 + ix^2), \qquad \xi^1\bar\xi^{0'} = \frac{1}{\sqrt 2}(x^1 - ix^2),$$

$$\xi^1\bar\xi^{1'} = \frac{1}{\sqrt 2}(x^0 - x^3)$$

and the condition imposed by equation (42) holds, which guarantees that it is a point on a null ray directed to the future ($x^0 > 0$).

Defining the spinors $\xi^A$ and $\eta^{A'}$ of rank 1 as complex vectors in a two-dimensional space, where $A, A' = 0,1$, and subject to the transformations

$$\xi_*^A = \alpha_B^A \xi^B, \qquad \eta_*^{A'} = \bar\alpha_{B'}^{A'}\eta^{B'}$$

where $\alpha_B^A$ and $\bar\alpha_{B'}^{A'}$ are complex conjugates with unit determinants $|\alpha_B^A| = |\bar\alpha_{B'}^{A'}| = 1$. We distinguish two classes of spinors: primed and unprimed, where the primed are subjected to complex conjugate transformations. As in tensor analysis, we can construct spinors with higher orders such as $\xi^{AB}$ or $\xi_C^{AB'}$ etc. with their appropriate transformations, e.g.,

$$\xi_*^{AB'} = \alpha_C^A \bar\alpha_{D'}^{B'}\xi^{CD'}.$$

If $\xi^A$ and $\eta^A$ are of the same class, then the determinant

$$\begin{vmatrix} \xi^0 & \xi^1 \\ \eta^0 & \eta^1 \end{vmatrix} = \xi^0\eta^1 - \xi^1\eta^0 \quad (44)$$

is invariant under unimodular transformations. Therefore, we may define a skew metric $\varepsilon_{AB}$ for the space such that $\varepsilon_{AB}\xi^A\eta^B$ is invariant. It follows from equation (44) that $\varepsilon_{00} = \varepsilon_{11} = 0$ and $\varepsilon_{01} = -\varepsilon_{10} = 1$. Rising and lowering the indices of spinors is done using $\varepsilon_{AB}$[10] and $\varepsilon^{AB}$.

As we represented $x^\mu$ in terms of two spinors $\xi^A$ and $\bar{\xi}^{A'}$ by

$$x^\mu \leftrightarrow \begin{pmatrix} \xi^0\bar{\xi}^{0'} & \xi^0\bar{\xi}^{1'} \\ \xi^1\bar{\xi}^{0'} & \xi^1\bar{\xi}^{1'} \end{pmatrix} \quad (45)$$

Generally, we can associate a 4-vector $X^\mu$ with a spinor of second rank $\xi^{AB'}$

$$X^\mu \leftrightarrow \begin{pmatrix} \xi^{00'} & \xi^{01'} \\ \xi^{10'} & \xi^{11'} \end{pmatrix} = X^{AB'}$$

Thus a 4-vector is associated with a Hermitian matrix. The invariance is now expressed by

$$g_{\mu\nu}X^\mu X^\nu = \varepsilon_{AC}\varepsilon_{B'D'}X^{AB'}X^{CD'}$$

Since a tetrad of the form $e^{AB'}$ belong to the complexification of the space[10], then there exists a scalar $\sigma_\mu^{AB'}$ such that $e^{AB'} = \sigma_\mu^{AB'}e^\mu$ or equivalently, $e^\mu = \sigma_{AB'}^\mu e^{AB'}$. The matrices $\sigma_{AB'}^\mu$ and $\sigma_\mu^{AB'}$ are Hermitian matrices obeying[11]

$$g_{\mu\nu} = \varepsilon_{AC}\varepsilon_{B'D'}\sigma_\mu^{AB'}\sigma_\nu^{CD'}$$

$$\sigma_{AB'}^\mu\sigma_\mu^{CD'} = \delta_C^A\delta_{D'}^{B'}, \qquad \sigma_{AB'}^\mu\sigma_\nu^{AB'} = \delta_\nu^\mu$$

$\sigma_i^{AB'}$ are connection symbols, called Infeld-van der Waerden symbols[10], any tensor with arbitrary rank can be related to its spinor equivalent by these symbols[1].

From equations (43) and (45), we find that

$$\sigma_{AB'}^0 = \frac{1}{\sqrt{2}}\begin{pmatrix} 1 & 0 \\ 0 & 1 \end{pmatrix}, \qquad \sigma_{AB'}^1 = \frac{1}{\sqrt{2}}\begin{pmatrix} 0 & 1 \\ 1 & 0 \end{pmatrix}, \qquad \sigma_{AB'}^2 = \frac{1}{\sqrt{2}}\begin{pmatrix} 0 & -i \\ i & 0 \end{pmatrix}, \qquad \sigma_{AB'}^3 = \frac{1}{\sqrt{2}}\begin{pmatrix} 1 & 0 \\ 0 & -1 \end{pmatrix}$$

Apart from the normalization factor, these are the Pauli matrices[11].

### 3-2-Dyad Formalism:

In this section we adopted the early capital Latin indices for spinor indices, early lowercase Latin letters in parentheses for dyad indices, and middle Greek letters for tensor indices.

Since the spacetime of general relativity is locally Minkowskian, we can setup at each point of spacetime an orthonormal dyad basis $\zeta_{(a)}^A$ and $\zeta_{(a')}^{A'}$ ($a,a'=0,1$; $A,A'=0,1$) for spinors as we setup an orthogonal tetrad basis $e_A^i$ ($A=0,1,2,3$; $i=0,1,2,3$) for tenors in a tetrad formalism.[11]

We define

---

[10] $\varepsilon_{AB}$ is a two-dimensional Levi-Civita symbol.

$$\zeta^A_{(0)} = o^A, \qquad \zeta^A_{(1)} = \iota^A$$

As in tetrad formalism, we can project any spinor $\xi_A$ onto the dyad basis, $\xi_{(a)} = \xi_A \zeta^A_{(a)}$, explicitly:

$$\xi_{(0)} = \xi_A o^A, \qquad \xi_{(1)} = \xi_A \iota^A$$

The spinors $o^A$ and $\iota^A$ and their complex conjugates determine the null vectors $l, n, m, \bar{m}$ by the correspondence:

$$l^\mu \leftrightarrow o^A \bar{o}^{B'}, \qquad m^\mu \leftrightarrow o^A \bar{\iota}^{B'}, \qquad \bar{m}^\mu \leftrightarrow \iota^A \bar{o}^{B'}, \qquad n^\mu \leftrightarrow \iota^A \bar{\iota}^{B'} \quad (46)$$

which satisfies the orthogonality conditions while the remaining products are zero. Thus, the dyad basis determines four null vectors which can be used as a basis for Newman-Penrose formalism.

Equation (46) yields the Infeld-van der Waerden symbols as

$$\left. \begin{aligned} l^\mu &= \sigma^\mu_{AB'} o^A \bar{o}^{B'} \\ m^\mu &= \sigma^\mu_{AB'} o^A \bar{\iota}^{B'} \\ \bar{m}^\mu &= \sigma^\mu_{AB'} \iota^A \bar{o}^{B'} \\ n^\mu &= \sigma^\mu_{AB'} \iota^A \bar{\iota}^{B'} \end{aligned} \right\} \Rightarrow \sigma^\mu_{AB'} = \frac{1}{\sqrt{2}} \begin{pmatrix} l^\mu & m^\mu \\ \bar{m}^\mu & n^\mu \end{pmatrix}$$

These are the generalization of Pauli matrices.

The directional derivatives of the Newman-Penrose formalism have the spinor equivalents:

$$\partial_{00'} = D, \qquad \partial_{11'} = \Delta, \qquad \partial_{01'} = \delta, \qquad \partial_{10'} = \bar{\delta}$$

Ricci rotation coefficients or the spin coefficients $\gamma_{ABC}$ are defined in the dyad basis by

$$\gamma_{(a)(b)(c)(d')} = \left[ \zeta_{(a)F} \right]_{;CD'} \zeta^F_{(b)} \bar{\zeta}^C_{(c)} \zeta^{D'}_{(d')}$$

The following table summarizes the notation[11]

| (a)(b) \ (c)(d') | 00 | 01 or 10 | 11 |
|---|---|---|---|
| 00' | $\kappa$ | $\varepsilon$ | $\pi$ |
| 10' | $\rho$ | $\alpha$ | $\lambda$ |
| 01' | $\sigma$ | $\beta$ | $\mu$ |
| 11' | $\tau$ | $\gamma$ | $\omega$ |

If we represent the 4-components Dirac spinor

$$\Psi = \begin{pmatrix} \psi_1 \\ \psi_2 \\ \psi_3 \\ \psi_3 \end{pmatrix}$$

in 2-components bispinor

$$\Psi = \begin{pmatrix} P_A \\ \bar{Q}_{B'} \end{pmatrix}$$

where $\bar{Q}_{B'}$ is complex conjugated. That yields Dirac equation in 2-spinor[13] or bispinor[11] formalism:

$$\sigma^i_{AB'} \partial_i P^A + i\mu_* \bar{Q}_{B'} = 0$$

$$\sigma^i_{AB'} \partial_i Q^A + i\mu_* \bar{P}_{B'} = 0$$

where $\mu_* = m/\sqrt{2}$ and $\sigma^i_{AB'}$ are Pauli matrices. This set of coupled differential equations are in Minkowski spacetime. In curved spacetime, ordinary derivatives are replaced by the covariant derivatives and Pauli matrices by Infeld-van der Wearden symbols.

$$\sigma^i_{AB'} P^A_{;i} + i\mu_* \bar{Q}^{C'} \varepsilon_{C'B'} = 0$$

$$\sigma^i_{AB'} Q^A_{;i} + i\mu_* \bar{P}^{C'} \varepsilon_{C'B'} = 0$$

Other references[10] uses the notation

$$\Psi = \begin{pmatrix} \psi_A \\ \varphi^{\dot{A}} \end{pmatrix}$$

and Dirac equation in Minkowski spacetime is expressed as

$$\partial_{A\dot{B}} \varphi^{\dot{B}} = \frac{m_0 c}{\hbar \sqrt{2}} \psi_A, \qquad \partial_{B\dot{A}} \psi^B = \frac{m_0 c}{\hbar \sqrt{2}} \varphi_{\dot{A}}$$

with $\partial_{A\dot{B}} = \frac{1}{\sqrt{2}} \sigma^i_{A\dot{B}} \partial_i$. And in curved spacetime as

$$\nabla_{A\dot{B}} \varphi^{\dot{B}} = \frac{m_0 c}{\hbar \sqrt{2}} \psi_A, \qquad \nabla_{B\dot{A}} \psi^B = \frac{m_0 c}{\hbar \sqrt{2}} \varphi_{\dot{A}}$$

Now we can get the final form of Dirac equation in Newman-Penrose formalism in chiral form[3] after assigning $F_1 = P^0, F_2 = P^1, G_1 = \bar{Q}^{1'}, G_2 = -\bar{Q}^{0'}$

$$(D + \varepsilon + \rho)F_1 + (\bar{\delta} + \pi + \alpha)F_2 = i\mu_* G_1$$

$$(\Delta + \mu - \gamma)F_2 + (\delta + \beta - \tau)F_1 = i\mu_* G_2$$

$$(D + \bar{\varepsilon} + \bar{\rho})G_2 - (\delta + \bar{\pi} - \bar{\alpha})G_1 = i\mu_* F_2$$

$$(\Delta + \bar{\mu} - \bar{\gamma})G_1 - (\bar{\delta} + \bar{\beta} - \bar{\tau})G_2 = i\mu_* F_1$$

We will adopt the notation used by Batic et al.[14]

$$(D + \varepsilon - \rho)P^0 + (\bar{\delta} + \pi - \alpha)P^1 = i\mu_* \bar{Q}^{1\prime}$$

$$(\delta + \beta - \tau)P^0 + (\Delta + \mu - \gamma)P^1 = i\mu_* \bar{Q}^{0\prime}$$

$$(D + \varepsilon - \rho)Q^{0\prime} + (\bar{\delta} + \pi - \alpha)P^{1\prime} = i\mu_* \bar{P}^1$$

$$(\delta + \beta - \tau)Q^{0\prime} + (\Delta + \mu - \gamma)Q^{1\prime} = i\mu_* \bar{P}^0$$

## IV-Non-existence of Bound States:

The problem of whether or not bound states exist for Dirac equation around a black hole is controversial. Finster et al.[15] showed that Dirac equation does not admit normalizable, time-periodic solutions in a non-extreme Reissner-Nordström black hole[11]. Finster et al.[16] derived a non-existence theorem for Kerr-Newman black hole[12]. Batic et al.[17] proved the non-existence of bound states in extreme Kerr-Newman metric[13]. Belgiorno et al[18] proved the non-existence of bound states of a Kerr-Newman-de Sitter black hole[14]. Schwarzschild black hole is the most controversial, since Zecca et al.[19] and Coatescu et al[20] showed that bound states exist for Fermions in Schwarzschild metric by approximating the radial system emerging from Dirac equation and on the construction of approximated solutions at the event horizon and far away from the black hole at space-like infinity, both of them ended up with an approximated spectrum resembling that of a hydrogen-like atom. However, Lasenby et al [21] showed that bound states cannot exist and only resonance is admitted. The problem with Zecca et al. and Coatescu et al. is that they oversimplify the system of differential equations governing the radial system.[14] Batic et al (2016)[14] approached is by reducing the radial system emerging from Dirac equation to a known differential equation, namely, the generalized Heun equation, despite the claim that the radial system of Dirac equation cannot be reduced to a known differential equation[6]. They solved the equation to find a condition for bound states solution, which can be satisfied if the energy is real . Then they approached the problem in an operator theoretic way, using the index theorem. In this work, the first approach is presented.

1-Ansatzes:

Recall the Schwarzschild metric

$$ds^2 = \left(1 - \frac{2M}{r}\right)dt^2 - \left(1 - \frac{2M}{r}\right)^{-1} dr^2 - r^2(d\theta^2 + sin^2\theta d\varphi^2)$$

where $r > 0, 0 \leq \theta \leq \pi, 0 \leq \varphi \leq 2\pi$.

By adopting Carter tetrad, the null tetrads are

$$l = \left(\frac{r}{\sqrt{2\Delta_r}}, -\frac{1}{r}\sqrt{\frac{\Delta_r}{2}}, 0, 0\right), \qquad n = \left(\frac{r}{\sqrt{2\Delta_r}}, \frac{1}{r}\sqrt{\frac{\Delta_r}{2}}, 0, 0\right), \qquad m = \left(0, 0, \frac{1}{r\sqrt{2}}, \frac{i}{r\sqrt{2}\sin\theta}\right)$$

With $\Delta_r = r^2 - 2Mr$

Using the following ansatz[11]

---

[11] Reissner-Nordström metric describes a static and charged black hole.
[12] Kerr-Newman metric describes a charged and rotating black hole.
[13] Extreme Kerr metric describes a rotating black hole with maximum angular momentum $a = M$
[14] Kerr-Newman-de Sitter is the Kerr-Newman metric with positive vacuum energy, i.e., a repulsive cosmological constant

$$P^0 = \frac{e^{i(\omega t + k\varphi)}}{\sqrt{r}\sqrt[4]{\Delta_r}} R_+(r) S_+(\theta), \qquad P^1 = \frac{e^{i(\omega t + k\varphi)}}{\sqrt{r}\sqrt[4]{\Delta_r}} R_-(r) S_-(\theta),$$

$$Q^{0\prime} = -\frac{e^{-i(\omega t + k\varphi)}}{\sqrt{r}\sqrt[4]{\Delta_r}} \overline{R_-}(r)\overline{S_+}(\theta), \qquad Q^{1\prime} = \frac{e^{-i(\omega t + k\varphi)}}{\sqrt{r}\sqrt[4]{\Delta_r}} \overline{R_+}(r)\overline{S_-}(\theta)$$

where $\omega$ is the energy of the particle and $k = \pm\frac{1}{2}, \pm\frac{3}{2}, \ldots$ is the azimuthal quantum number.

This procedure leads to the following radial and angular systems

$$\begin{pmatrix} \sqrt{\Delta_r}D_0 & -(\lambda + imr) \\ -(\lambda - imr) & \sqrt{\Delta_r}D_0^\dagger \end{pmatrix}\begin{pmatrix} R_- \\ R_+ \end{pmatrix} = 0, \qquad \begin{pmatrix} \mathcal{L}^{\frac{1}{2}} & -\lambda \\ \lambda & \mathcal{L}_{\frac{1}{2}}^\dagger \end{pmatrix}\begin{pmatrix} S_- \\ S_+ \end{pmatrix} = 0$$

where $\lambda$ is a constant of separation[11] and the differential operators

$$D_0 = \frac{d}{dr} + i\frac{\omega r^2}{\Delta_r}, \qquad D_0^\dagger = \frac{d}{dr} - i\frac{\omega r^2}{\Delta_r}, \qquad \mathcal{L}^{\frac{1}{2}} = \frac{d}{d\theta} + \frac{k}{\sin\theta} + \frac{1}{2}\cot\theta,$$

$$\mathcal{L}_{\frac{1}{2}}^\dagger = \frac{d}{d\theta} - \frac{k}{\sin\theta} + \frac{1}{2}\cot\theta$$

With the following property of radial spinors

$$\overline{R_-}(r) = R_+(r), \qquad \overline{R_+}(r) = R_-(r)$$

we can write the radial system as

$$\left.\begin{aligned} \frac{dR_-}{dr} + i\frac{\omega r}{r - 2M} R_- &= \frac{\lambda + imr}{\sqrt{r^2 - 2Mr}} R_+ \\ \frac{dR_+}{dr} - i\frac{\omega r}{r - 2M} R_+ &= \frac{\lambda - imr}{\sqrt{r^2 - 2Mr}} R_- \end{aligned}\right\} \qquad (46)$$

and the solution of Dirac equation is

$$P^0 = \frac{e^{i(\omega t + k\varphi)}}{\sqrt{r}\sqrt[4]{\Delta_r}} R_+(r) S_+(\theta), \qquad P^1 = \frac{e^{i(\omega t + k\varphi)}}{\sqrt{r}\sqrt[4]{\Delta_r}} R_-(r) S_-(\theta),$$

$$Q^{0\prime} = -\frac{e^{-i(\omega t + k\varphi)}}{\sqrt{r}\sqrt[4]{\Delta_r}} R_+(r)\overline{S_+}(\theta), \qquad Q^{1\prime} = \frac{e^{-i(\omega t + k\varphi)}}{\sqrt{r}\sqrt[4]{\Delta_r}} R_-(r)\overline{S_-}(\theta)$$

If we let

$$P^0 = \frac{F_1}{\sqrt{r}\sqrt[4]{\Delta_r}}, \qquad P^1 = \frac{F_2}{\sqrt{r}\sqrt[4]{\Delta_r}}, \qquad Q^{0\prime} = -\frac{\overline{G_2}}{\sqrt{r}\sqrt[4]{\Delta_r}}, \qquad Q^{0\prime} = \frac{\overline{G_1}}{\sqrt{r}\sqrt[4]{\Delta_r}}$$

Dirac equation can be written as

$$(\mathcal{R} + \mathcal{A})\Psi = 0; \ \ \Psi = \begin{pmatrix} F_1 \\ F_2 \\ G_1 \\ G_2 \end{pmatrix} \qquad (47)$$

with

$$\mathcal{R} = \begin{pmatrix} -imr & 0 & \sqrt{\Delta_r}D_+ & 0 \\ 0 & imr & 0 & -\sqrt{\Delta_r}D_- \\ -\sqrt{\Delta_r}D_- & 0 & -imr & 0 \\ 0 & \sqrt{\Delta_r}D_+ & 0 & -imr \end{pmatrix}, \quad \mathcal{A} = \begin{pmatrix} 0 & 0 & 0 & \mathcal{L}_- \\ 0 & 0 & -\mathcal{L}_+ & 0 \\ 0 & \mathcal{L}_- & 0 & 0 \\ \mathcal{L}_+ & 0 & 0 & 0 \end{pmatrix} \quad (48)$$

with the differential operators,

$$D_\pm = \partial_r \pm \frac{r^2}{\Delta_r}\partial_t, \qquad \mathcal{L}_\pm = \partial_\theta \pm \frac{i}{\sin\theta}\partial_\varphi + \frac{1}{2}\cot\theta \qquad (49)$$

From equation (47) we can construct Dirac equation in the Hamiltonian form:

$$i\partial_t\Psi = \mathcal{H}\Psi$$

with the Hamiltonian from equations (48) and (49)

$$\mathcal{H} = \frac{i\Delta_r}{r^2}\begin{pmatrix} -\partial_r & 0 & 0 & 0 \\ 0 & \partial_r & 0 & 0 \\ 0 & 0 & \partial_r & 0 \\ 0 & 0 & 0 & -\partial_r \end{pmatrix} - \frac{m}{r}\sqrt{\Delta_r}\begin{pmatrix} 0 & 0 & 1 & 0 \\ 0 & 0 & 0 & 1 \\ 1 & 0 & 0 & 0 \\ 0 & 1 & 0 & 0 \end{pmatrix} + i\frac{\sqrt{\Delta_r}}{r^2}\begin{pmatrix} 0 & \mathcal{L}_- & 0 & 0 \\ \mathcal{L}_+ & 0 & 0 & 0 \\ 0 & 0 & 0 & -\mathcal{L}_- \\ 0 & 0 & -\mathcal{L}_+ & 0 \end{pmatrix}$$

After the introduction of the Hamiltonian, the following inner product is defined outside the black hole for $r > 2M$ on hypersurfaces of constant time:

$$\langle\Psi|\Phi\rangle = \int\limits_{2M}^{+\infty} dr\frac{r^2}{\Delta_r}\int\limits_0^\pi d\theta\sin\theta\int\limits_0^{2\pi} d\varphi\,\Psi^\dagger\Phi$$

The Hamiltonian $\mathcal{H}$ is symmetric and self-adjoint with respect to this inner product[14]. So, a time-periodic solution of the form

$$\Psi(t,r,\theta,\varphi) = e^{-i\omega t}\Psi_0(r,\theta,\varphi); \ \ \omega \in \mathbb{R}, \Psi_0 \neq 0$$

Such solution is a bound state if it is normalizable: $\langle\Psi|\Psi\rangle = \langle\Psi_0|\Psi_0\rangle < \infty$. In what follows, the conditions for normalization of radial and angular systems will be derived, from which the non-existence of bound states is concluded.

By introducing Chandrasekhar ansatz

$$\Psi_0(r,\theta,\varphi) = e^{ik\varphi}\begin{pmatrix} R_+(r)S_+(\theta) \\ R_-(r)S_-(\theta) \\ R_-(r)S_+(\theta) \\ R_-(r)S_-(\theta) \end{pmatrix}$$

for energy $\omega \in \mathbb{R}$ and a constant of separation $\lambda \in \mathbb{R}$, the non-trivial solutions for the radial and angular systems are

$$R(r) = \begin{pmatrix} R_-(r) \\ R_+(r) \end{pmatrix}, \qquad S(\theta) = \begin{pmatrix} S_-(\theta) \\ S_+(\theta) \end{pmatrix}$$

satisfying the normalizable conditions respectively

$$\int_{2M}^{+\infty} dr\, \frac{r^2}{\Delta_r} |R(r)|^2 < \infty, \qquad \int_0^{\pi} d\theta \sin\theta\, |S(\theta)|^2 < \infty \quad (50)$$

## 2-Angular system:

Batic et al. (2004)[22] studied the eigenvalue problem of an operator $A(\kappa;\mu,\nu)$, for a given azimuthal quantum number $\kappa = k - \frac{1}{2}$, of Chandrasekhar-Page angular equation arising from the separation of Dirac equation in Kerr-Newman geometry in terms of the parameters $\mu \coloneqq am$ and $\nu \coloneqq a\omega$, where the Kerr parameter $a$ is the angular momentum per unit mass of a black hole, $m$ is the mass of the particle, $\omega$ is the energy.

The Chandrasekhar-Page equations in Kerr-Newman geometry are

$$\mathcal{L}_{\frac{1}{2}}^+ S_{+\frac{1}{2}} = (am\cos\theta - \lambda)S_{-\frac{1}{2}}$$

$$\mathcal{L}_{\frac{1}{2}}^- S_{-\frac{1}{2}} = (am\cos\theta + \lambda)S_{+\frac{1}{2}}$$

where,

$$\mathcal{L}_{\frac{1}{2}}^{\pm} = \partial_\theta \pm Q(\theta) + \frac{\cot\theta}{2};\ \ Q(\theta) \coloneqq a\omega\sin\theta + \frac{\kappa}{\sin\theta}\,, \theta \in (0,\pi)$$

For the case of $\mu,\nu{=}0$, a nontrivial solution $S(\theta)$ satisfies

$$\int_0^{\pi} d\theta\, |S(\theta)|^2 < \infty \quad (51)$$

and the operator $A(\kappa;0,0)$ has discrete eigenvalues given by $\{\lambda_n^{\pm}: n = 0,1,2\dots\}$

We can conclude from equation (51) that the second equation in (50) is satisfied, and the result is discrete eigenvalues for the angular system arising from Dirac equation in Schwarzschild geometry.

## 3-Radial system:

The approach in this section is to derive a set of necessary and sufficient conditions for the existence of bound states. It follows that this set of conditions cannot be satisfied if $\omega$ is real. If $\omega$ is allowed to be complex, the solutions are resonance states instead of bounded.

Suppose $\Omega = 2M\omega$, $\mu = 2Mm$, and the rescaled variable $\rho = r/2M$ with $\rho \in (1, \infty)$. Let

$$f(\rho) = \frac{\rho}{\rho - 1} \ , \ \ g(\rho) = \frac{\lambda + i\mu\rho}{\sqrt{\rho^2 - \rho}}$$

The radial system of equation (46) can be written as

$$\frac{dR_-(\rho)}{d\rho} + i\Omega f R_-(\rho) = g R_+(\rho), \qquad \frac{dR_+(\rho)}{d\rho} - i\Omega f R_+(\rho) = \bar{g} R_-(\rho) \quad (52)$$

Equation (52) decouples into two linear differential equations of second order

$$\left.\begin{array}{l} R_-'' - \dfrac{g'}{g} R_-' + \left[\Omega^2 f^2 - |g|^2 + i\Omega\left(f' - f\dfrac{g'}{g}\right)\right] R_- = 0 \\[3mm] R_+'' - \dfrac{\bar{g}'}{\bar{g}} R_+' + \left[\Omega^2 f^2 - |g|^2 - i\Omega\left(f' - f\dfrac{\bar{g}'}{\bar{g}}\right)\right] R_+ = 0 \end{array}\right\} \quad (53)$$

By making a partial fraction expansion of the coefficient functions, equations (53) become in the form of generalized Heun equation

$$R_\pm'' + p_\pm(\rho) R_\pm' + q_\pm(\rho) R_\pm = 0 \quad (54)$$

with

$$p_\pm(\rho) = \frac{1/2}{\rho} + \frac{1/2}{\rho - 1} - \frac{1}{\rho - c_\pm} \ , \qquad c_\pm = \mp i\frac{\lambda}{\mu}$$

$$q_\pm(\rho) = \Omega^2 - \mu^2 + \frac{\lambda^2}{\rho} + \frac{\alpha_\pm}{\rho - 1} + \frac{\beta_\pm}{(\rho - 1)^2} + \frac{\gamma_\pm}{\rho - c_\pm}$$

where,

$$\alpha_\pm = 2\Omega^2 - \lambda^2 - \mu^2 - \gamma_\pm, \qquad \beta_\pm = \Omega^2 \pm i\frac{\Omega}{2}, \qquad \gamma_\pm = \mp\frac{\lambda\Omega}{i\lambda \pm \mu}$$

By the transformation

$$R_\pm = e^{-\sqrt{\mu^2 - \Omega^2}\,\rho} \tilde{R}_\pm$$

equation (54) becomes

$$\tilde{R}_\pm'' + P_\pm(\rho) \tilde{R}_\pm' + Q_\pm(\rho) \tilde{R}_\pm = 0 \quad (55)$$

with

$$P_\pm(\rho) = p_\pm(\rho) - 2\sqrt{\mu^2 - \Omega^2}, \qquad Q_\pm(\rho) = \frac{\sigma}{\rho} + \frac{\tilde{\alpha}_\pm}{\rho - 1} + \frac{\beta_\pm}{(\rho - 1)^2} + \frac{\tilde{\gamma}_\pm}{\rho - c_\pm}$$

where,

$$\sigma = \lambda^2 - \frac{\sqrt{\mu^2 - \Omega^2}}{2}, \qquad \tilde{\alpha}_\pm = \alpha_\pm - \frac{\sqrt{\mu^2 - \Omega^2}}{2}, \qquad \tilde{\gamma}_\pm = \gamma_\pm + \sqrt{\mu^2 - \Omega^2}$$

The transformation

$$\bar{R}_\pm = (\rho - 1)^{\delta_\pm} \hat{R}_\pm$$

eliminates the $1/(\rho - 1)^2$ term in $Q_\pm(\rho)$, with $\delta_\pm \in \mathbb{C}$. Hence $\hat{R}_\pm$ satisfies the differential equation

$$\hat{R}''_\pm + \mathfrak{p}_\pm(\rho)\hat{R}'_\pm + \mathfrak{q}_\pm(\rho)\hat{R}_\pm = 0 \qquad (56)$$

with

$$\mathfrak{p}_\pm(\rho) = \frac{1/2}{\rho} + \frac{2\delta_\pm + 1/2}{\rho - 1} - \frac{1}{\rho - c_\pm} - 2\sqrt{\mu^2 - \Omega^2}$$

$$\mathfrak{q}_\pm(\rho) = \frac{\hat{\sigma}_\pm}{\rho} + \frac{\hat{\alpha}_\pm}{\rho - 1} + \frac{\hat{\beta}_\pm}{(\rho - 1)^2} + \frac{\hat{\gamma}_\pm}{\rho - c_\pm}$$

where,

$$\hat{\sigma}_\pm = \sigma - \frac{\delta_\pm}{2}, \qquad \hat{\alpha}_\pm = \tilde{\alpha}_\pm - 2\delta_\pm\sqrt{\mu^2 - \Omega^2} + \frac{\delta_\pm}{2} + \frac{\delta_\pm}{c_\pm - 1}$$

$$\hat{\beta}_\pm = {\delta_\pm}^2 - \frac{\delta_\pm}{2} + \beta_\pm, \qquad \hat{\gamma}_\pm = \tilde{\gamma}_\pm - \frac{\delta_\pm}{c_\pm - 1}$$

Notice that $\hat{\beta}_\pm = 0$ whenever $\delta_\pm = \pm i\Omega$.

Therefore, the radial system described by equations (46) decouples into the following generalized Heun equations

$$\hat{R}''_\pm + \left( \sum_{n=0}^{2} \frac{1 - \mu_{n,\pm}}{\rho - \rho_n} + \alpha \right) \hat{R}'_\pm + \frac{\beta_{0,\pm} + \beta_{1,\pm}\rho + \beta_{2,\pm}\rho^2}{\prod_{n=0}^{2}(\rho - \rho_n)} \hat{R}_\pm = 0 \qquad (57)$$

where $\alpha = -2\sqrt{\mu^2 - \Omega^2}$.

The exponents associated with the simple singularities $\rho_0 = 0$, $\rho_1 = 1$ and $\rho_2 = c_\pm = \mp i\frac{\lambda}{\mu}$ are respectively $\mu_{0,\pm} = \frac{1}{2}$, $\mu_{1,\pm} = \frac{1}{2} \mp 2i\Omega$ and $\mu_{2,\pm} = 2$. With,

$$\beta_{0,\pm} = c_\pm\hat{\sigma}_\pm, \qquad \beta_{1,\pm} = -c_\pm(\hat{\alpha}_\pm + \hat{\sigma}_\pm) - \hat{\gamma}_\pm, \qquad \beta_{2,\pm} = \hat{\alpha}_\pm + \hat{\sigma}_\pm + \hat{\gamma}_\pm$$

and

$$\hat{\alpha}_\pm = 2\Omega^2 - \lambda^2 - \mu^2 - \frac{\sqrt{\mu^2 - \Omega^2}}{2}(1 \pm 4i\Omega) \mp i\frac{\Omega}{2}, \qquad \hat{\gamma}_\pm = \pm i\Omega + \sqrt{\mu^2 - \Omega^2},$$

$$\hat{\sigma}_\pm = \lambda^2 - \frac{\sqrt{\mu^2 - \Omega^2}}{2} \mp i\frac{\Omega}{2}$$

The problem of the existence of bound states is now reduced to whether or not the generalized Heun equation admits polynomial solutions for $|\Omega| < \mu$.

Let us write equation (57) as

$$\rho(\rho - 1)(\rho - c_\pm)\hat{R}''_\pm + \mathfrak{B}_\pm(\rho)\hat{R}'_\pm + \mathfrak{D}_\pm(\rho)\hat{R}_\pm = 0 \quad (58)$$

with

$$\mathfrak{B}_\pm(\rho) = (1 - \mu_{0,\pm})(\rho - 1)(\rho - c_\pm) + (1 - \mu_{1,\pm})\rho(\rho - c_\pm) + (1 - \mu_{2,\pm})\rho(\rho - 1)$$
$$+ \alpha\rho(\rho - 1)(\rho - c_\pm)$$

$$\mathfrak{D}_\pm(\rho) = \beta_{0,\pm} + \beta_{1,\pm}\rho + \beta_{2,\pm}\rho^2$$

Suppose that

$$\hat{R}_\pm(\rho) = \sum_{n=0}^{\infty} d_{n,\pm}(\rho - 1)^n \quad (59)$$

In order for this infinite series to stop at some value $n$, first, we need to derive the recurrence relation satisfied by the coefficient $d_{n,\pm}$. If we let $\tau = \rho - 1$, equation (58) becomes

$$\mathfrak{A}(\tau)\hat{R}_\pm(\tau) + \mathfrak{B}_\pm(\tau)\hat{R}'_\pm(\tau) + \mathfrak{D}_\pm(\tau)\hat{R}_\pm(\tau) = 0 \quad (60)$$

with

$$\mathfrak{A}(\tau) = \tau^3 + a_{1,\pm}\tau^2 + a_{2,\pm}\tau, \qquad \mathfrak{B}_\pm(\tau) = \alpha\tau^3 + b_{1,\pm}\tau^2 + b_{2,\pm}\tau + b_{3,\pm},$$

$$\mathfrak{D}_\pm(\tau) = \beta_{2,\pm}\tau^2 + c_{1,\pm}\tau + c_{2,\pm}$$

where

$$a_{1,\pm} = 2 - c_\pm, \qquad a_{2,\pm} = 1 - c_\pm, \qquad b_{1,\pm} = \alpha(2 - c_\pm) \pm 2i\Omega$$

$$b_{2,\pm} = \alpha(1 - c_\pm) - c_\pm(1 \pm 2i\Omega) + \frac{1}{2} \pm 4i\Omega, \qquad b_{3,\pm} = (1 - c_\pm)\left(\frac{1}{2} \pm 2i\Omega\right)$$

$$\beta_{2,\pm} = 2\Omega^2 - \mu^2 \mp 2i\Omega\sqrt{\mu^2 - \Omega^2}, \qquad c_{1,\pm} = \beta_{1,\pm} + 2\beta_{2,\pm}, \qquad c_{2,\pm} = \beta_{0,\pm} + \beta_{1,\pm} + \beta_{2,\pm}$$

Substituting equation (59) in equation (60) and shifting indices appropriately yield the four-term recurrence relation

$$\varphi_1(n+1)d_{n+1,\pm} + \varphi_2(n)d_{n,\pm} + \varphi_3(n-1)d_{n-1,\pm} + \varphi_4(n-2)d_{n-2,\pm} = 0 \quad , n \in \mathbb{N} \quad (61)$$

with

$$\varphi_1(m) = a_{2,\pm} m(m-1) + b_{3,\pm} m, \qquad \varphi_2(m) = a_{1,\pm} m(m-1) + b_{2,\pm} m + c_{2,\pm}$$

$$\varphi_3(m) = m(m-1) + b_{1,\pm} m + c_{1,\pm}, \qquad \varphi_4(m) = \alpha m + \beta_{2,\pm}$$

Let $n = N + 2$ in equation (16) to find the conditions for which we may have polynomial solutions of the form

$$\sum_{n=0}^{N} d_{n,\pm} \tau^n, \qquad N = 0,1,2,\dots$$

Then equation (61) becomes

$$\varphi_1(N+3) d_{N+3,\pm} + \varphi_2(N+2) d_{N+2,\pm} + \varphi_3(N+1) d_{N+1,\pm} + \varphi_4(N) d_{N,\pm} = 0, \qquad N \in \mathbb{N}$$

For any fixed $N$ we will have a polynomial solution of degree $N$ whenever

$$\varphi_4(N) = 0, \qquad d_{N+1,\pm} = d_{N+2,\pm} = 0$$

This implies that $\alpha N + \beta_{2,\pm} = 0$ with $\alpha = -2\sqrt{\mu^2 - \Omega^2}$. Bound states solutions exist for real values of $\Omega$ satisfying simultaneously the above set of equations and such that $|\Omega| < \mu$. The condition for $\varphi_4(N) = 0$ reads

$$2\Omega^2 - \mu^2 - 2N\sqrt{\mu^2 - \Omega^2} = \pm 2i\Omega\sqrt{\mu^2 - \Omega^2} \qquad (62)$$

If we suppose that there exists a bound state of energy $\Omega_N \in \mathbb{R}$ for some $N \in \mathbb{N}$, then the L.H.S of equation (62) represents a real number and the R.H.S represents an imaginary number. Therefore, by contradiction, no polynomial solution exists for any $N \in \mathbb{N}$, and hence no bound states exist.

*"Secretly, bring It pure*

*Publicly, bring It mixed"*

*Diwan Ayuha al-Saki. Sheikh Abdul Rahman Al-Shaghouri*

**V-Conclusion:**

We concluded that no Fermionic bound state exists in the background of Schwarzschild metric for energy $\omega < m$[15]. A. Lasenby et al.[21] studied the full atlas including the inside of the black hole, concluding that the singularity acts like a current sink. This means that the normalizable states must decay in time, therefore the energy eigenvalues are complex and considered as resonance states. Though we neglected the singularity, we got the same conclusion where the horizon is sufficient to explain this phenomenon. Resonant states in quantum mechanics are related to two quantities[23], the decrease of the number of particles if it is counted for a fixed volume and the momentum flux going out of the system. As mentioned, the singularity acts like a sink, in addition to that, the physical interpretation of these states is that they create an outward flux of particles at the horizon. This flux is described by Fermi-Dirac statistics at a Hawking temperature $T = 1/8\pi M k_B$. Surprisingly, this distribution is obtained without the apparatus of quantum field theory. Even if we consider particles of integer spin, it yields the same temperature with a flux described by Bose-Einstein distribution.[6] If we model the vacuum in terms of a Dirac's sea of negative energy states having the same decay factor, we conclude that the vacuum itself is decaying and could contribute to Hawking radiation.[21]

To elaborate on the meaning of this finding, let us consider some analogies between the classical cases and the quantum cases. In classical electrodynamics, a particle orbiting a point source should radiate, and by Gauss law, the particle cannot tell whether the source has a structure, or it is point-like. This problem is resolved in quantum mechanics by finding stable bound states, and describing the particle by a wave function defined over the whole space. This means that the particle can tell if the source has a structure or not. On the other hand, a body can orbit a black hole on a geodesic outside the horizon with a stable orbit. In our case, the Fermion does not have a stable orbit around the black hole even at a classical distance, and it is not supposed to 'feel' the singularity beyond the horizon, because Schwarzschild metric describes a static massive body, either it is a black hole or a star.

This demonstrates that our current understanding of the interaction between gravity and quantum theory is limited. And a complete treatment for the decay onto the singularity, with the fact that there is no emission associated with such transition, requires a quantum theory of the singularity, which does not exist yet. Batic et al[14] suggested that such behavior arises from the non-local nature of quantum mechanics. Based on the fact that in the Planck regime, the quantum fluctuations become so strong that the classical description breaks down, Ashtekar and Bojowald[24] suggested that instead of introducing new principles or requiring *ad hoc* assumptions, one can extend the spacetime continuum of general relativity to a discrete quantum geometry. It is important to clarify the type of problems encountered here. We are considering a perturbative approach[25] which is also known as 'semi-classical quantum gravity', referring to set a well-defined classical background geometry on which the quantum fluctuations are occurring. In contrast to the 'low energy effective field theoretic' approach dealing with the nature of singularities, Cauchy horizons, unification of gravity with other forces etc.

---

[15] Batic et al. (2008) proved the non-existence of Fermionic bound states for any real $\omega$ using the index theorem.